\documentclass[aps,prb,superscriptaddress,showpacs]{revtex4}
\usepackage{amssymb}
\usepackage{amsmath}
\usepackage{graphicx}

\begin{document}

\title{Effects of anisotropic elasticity in the problem of domain formation
and stability of monodomain state in ferroelectric films}
\author{A.M. Bratkovsky}
\affiliation{Hewlett-Packard Laboratories, 1501 Page Mill Road, Palo Alto, California
94304}
\author{A.P. Levanyuk}
\affiliation{Hewlett-Packard Laboratories, 1501 Page Mill Road, Palo Alto, California
94304}
\affiliation{Dept. Fiz. Mat. Cond., Universidad Autonoma de Madrid, Madrid 28049, Spain}
\affiliation{Moscow Institute of Radioengineering, Electronics and Automation, Moscow
117454, Russia}
\date{January 1, 2011 }

\begin{abstract}
We study cubic ferroelectrics films that become uniaxial with a polar axis
perpendicular to the film because of a misfit strain due to a substrate. The
main present result is the analytical account for the elastic anisotropy as
well as the anisotropy of the electrostriction. They define, in particular,
an orientation of the domain boundaries and stabilizing or destabilizing
effect of inhomogeneous elastic strains on the single domain state. We apply
the general results to perovskite systems like BaTiO$_{3}$/SrRuO$_{3}$/SrTiO$%
_{3}$ films and find that at least not far from the ferroelectric phase
transition the equilibrium domain structure consists of the stripes along
the cubic axes or at 45 degrees to them. We have also showed that in this
system the inhomogeneous strains increase stability with regards to the
small fluctuations of the metastable single domain state, which may exist
not very close to the ferroelectric transition. The latter analytical result
is in qualitative agreement with the numerical result by Pertsev and
Kohlstedt [Phys. Rev. Lett. \textbf{98}, 257603 (2007)], but we show that
the effect is much smaller than those authors claim. We have found also that
under certain conditions on the material constants, which are not satisfied
in the perovskites but are not forbidden either, a checkerboard domain
structure can be realized instead of the stripe-like one and that the
polarization-strain coupling decreases stability of a single domain state
instead of increasing it. The single domain state is metastable at certain
large thicknesses and becomes suitable for memory applications at even
larger thicknesses when the lifetime of the metastable state becomes
sufficiently large.
\end{abstract}

\maketitle

\section{Introduction}

Properties of domain structures in thin ferroelectric films is currently a
focus of extensive research. It is expected, quite naturally, that an
understanding and an ability to control these properties will determine the
prospects of applications of nanometer-size ferroelectrics. It depends
critically on the external conditions like presence or absence of
electrodes. In this paper, we discuss domain structures in a system, which
is, perhaps, the most important for applications: a ferroelectric film with
electrodes. The polar axis of the material is perpendicular to the film
plane and the electrodes are `real' meaning that the electric field
penetrates into them, although only over tiny depths $<1${$\mathring{A}$}.
This is an adequate model for the perovskite ferroelectric films on a
substrate with compressive strain, like BaTiO$_{3}$/SrRuO$_{3}$/SrTiO$_{3}$
(BTO/SRO/STO)\cite{KimL05,KimAPL05,nohNodl06,ghosez03,BLapl06} where the
misfit strain drives the FE\ film into a uniaxial state. We supplement our
analytical results with the relevant numerical estimates for BaTiO$_{3}$\
(BTO), PbTiO$_{3}$ (PTO), and Pb(Zr$_{0.5}$Ti$_{0.5}$)O$_{3}$ (PZT) using
the material constants available in the literature

Incomplete screening of the depolarizing field by SrRuO$_{3}$ electrode
leads to an absolute instability of a single domain state and formation of a
sinusoidal domain structure when thickness of BaTiO$_{3}$ film is slightly
above the minimal thickness compatible with ferroelectricity in this system
\cite{BLapl06, BLreview, Junquera08}. It seems that this situation is
typical of real electrodes and we shall consider this case only. To find the
minimal thickness, one does not need to take into account higher order terms
in the Landau-Ginzburg-Devonshire (LGD) free energy including the terms
describing the electrostriction since the problem of \emph{stability} of the
paraelectric phase is \emph{linear}\cite{ChT82}. But in order to reveal the
characteristics of this structure, i.e. to find out if the equilibrium
structure is stripe-like or checkerboard and how the domain boundaries are
oriented one has to take into account the anizotropy of elastic and
electrostrictive properties of the ferroelectric. This is the main goal of
the present paper. Specifically, we consider the case of cubic crystal
anisotropy of elastic and electrostrictive properties only. This is relevant
for films of cubic perovskites which become tetragonal because of in-plane
misfit strains due to cubic substrates like in the above-mentioned system.
The change of cubic anisotropy to tetragonal affects most strongly the
dielectric properties since the crystals are \textquotedblleft soft"
dielectrically. They have much smaller effect on the elastic and
electrostrictive properties, which can be considered to be the same as in
cubic parent crystals.

Explicit account for the electrostriction and anisotropic elasticity is
relevant also for study of stability of single domain state. This has been
correctly pointed out by Pertsev and Kohlstedt \cite{perkohl06,perkohl07}
although these authors have missed several important points. Importantly,
however \cite{BLcomPer07}, they overlooked that the state whose stability
they were studying was actually \emph{metastable}. Therefore, its stability
with respect to small fluctuations did not mean that this state can be used
in memory applications. Indeed, its lifetime is very short if the film
thickness is not sufficiently larger than that calculated by Pertsev and
Kohlstedt as the limit of single domain stability. We shall also discuss
this stability among other questions This makes sense because of several
reasons. \emph{First}, Pertsev and Kohlstedt performed numerical
calculations using material constants for BaTiO$_{3}$ and Pb(Zr$_{0.5}$Ti$%
_{0.5}$)O$_{3}$ and the electrode parameters of SrRuO$_{3}$ while our
results are analytical and apply to any material of the same symmetry.
Moreover, our method applies to other symmetries as well. \emph{Second},
Pertsev and Kohlstedt studied stability of the single domain state with
respect to \textquotedblleft polarization wave"-like fluctuations with a
single specific direction of the wave vector, while we consider waves with $%
\boldsymbol{k}$-vectors in arbitrary direction. \emph{Third}, Pertsev and
Kohlstedt apparently misinterpreted their own results by mixing together the
well-known effect of \emph{homogeneous} misfit strains and the effect of
strains due to \emph{inhomogeneous} polarization. In fact, the misfit strain
simply results in renormalization of the materials constants and was
effectively taken into account by all the previous authors. Only the account
of the inhomogeneous polarization was pioneered in Ref.\cite{perkohl07}. We
show that this effect was vastly overestimated by Pertsev and Kohlstedt. In
fact, the formal difference by more than an order of magnitude between the
results with and without account for electrostriction that they claimed
stems from improper comparison of the compressed film with the materials
constants renormalized by the misfit strain to one without \emph{any} such
renormalization at all, and \emph{not} from the effect of the inhomogeneous
strains on stability\ of single domain state.

Our analytical calculations provide a general view on the role of the
inhomogeneous strains in stability of single domain state. In particular,
they reveal a possibility which seems academic at the moment but no reason
is seen to exclude it altogether. We mean a specific state where \emph{%
elasticity} provokes domain formation of \emph{non-ferroelastic}
ferroelectric $180^\circ$ domains. Such a state is realized if a certain
condition on the electrostrictive and elastic constants is met. We are not
aware of an experimental realization of these conditions but we cannot find
arguments prohibiting them. It is worth mentioning that a qualitative
conclusion about possibility of both stabilizing and destabilizing role of
inhomogeneous elastic strains for single domain state in ferroelectric films
on substrates has been made in our previous paper where we considered an
academic case of a\ single electrostriction constant and assumed isotropic
elasticity \cite{BLKhach}. A surprising result of the present work is that
the \emph{destabilizing} effect of the inhomogeneous strains may be very
large contrary to the stabilizing one. Another unexpected result is the
possibility of a checkerboard domain state if some conditions on the
material constants are met. Let us mention that without account for the
anisotropic polarization-strain coupling one comes to the conclusion about
impossibility of such a state. For the perovskites this conclusion remains
valid but not in the general case.

Studying the sinusoidal domain structure in BTO, PTO and PZT films on SrTiO$%
_{3},$we find that the equilibrium orientation of the "domain walls" is
parallel $\left( \text{perpendicular}\right) $ to the cubic axes in the film
plane for BTO and PTO and is at $45^\circ$ to these axes for PZT. In all
cases the free energy of the sinusoidal domain structure depends very weakly
on the domain wall orientation. This is mainly due to both systems being
nearly isotropic elastically and, additionally, the relevant
electrostriction constant is relatively small. This observation may be
important for understanding domain creation at smallest thicknesses of the
ferroelectric films. For BTO/SRO/STO system, our analytical calculations
provide confirmation of the qualitative result of Pertsev and Kohlstedt
about stabilizing effect of inhomogeneous elastic strains for single domain
state \emph{in this system }but with the above mentioned strong disagreement
with their statement about importance of this effect.

Having mentioned advantages and new possibilities provided by analytical
calculations we should mention also their inherent shortcomings. Our
analytical method is feasible within a certain approximation only. This
approximation implies that the domain period is less than the film
thickness. This condition is fulfilled for thick enough films but in very
thin films the two quantities are in fact comparable. Therefore, the
accuracy of our calculations should be investigated for these films, so the
new numerical studies are desirable. We do not expect, however, that the
difference between the results of approximated and more exact calculations
either within a continuous medium theory or within microscopic theories will
be very large given close results of continuous and first principles
theories even for films that are just several unit cell thick (see, e.g.,
\cite{Junquera08}).

The paper is organized as follows: we describe the approximations used and
define the terms in the LGD free energy that can be neglected within our
approximation in Sec.\ref{sec:outline}. This let us avoid unnecessary
lengthy formulas in the rest of the paper. We spell out the constituent
equations in Sec.\ref{sec:constit} \ and then solve the general problem for
the `polarization wave' (embryonic stripe domains)\ in the FE film with full
account for elastic coupling. This is further used in Sec.\ref{sec:orient}
to determine that the domain walls align with crystallographic cubic axes.
Then, we find the conditions when the monodomain state first loses its
stability with regards to the stripe domain structure in Sec.\ref{sec:mdloss}%
. One previously unexplored possibility is that the system can lose
stability with regards to checkerboard domain structure, but our results in
Sec.\ref{sec:checker} show that such a structure is absolutely unstablein
perovskites although it is not necessarily so in the general case. We
summarize the present results in the Conclusions.

\section{Outline of the method and the approximations used\label{sec:outline}%
}

\begin{figure}[tbp]
\begin{center}
\includegraphics[angle=0,width=0.80\textwidth]{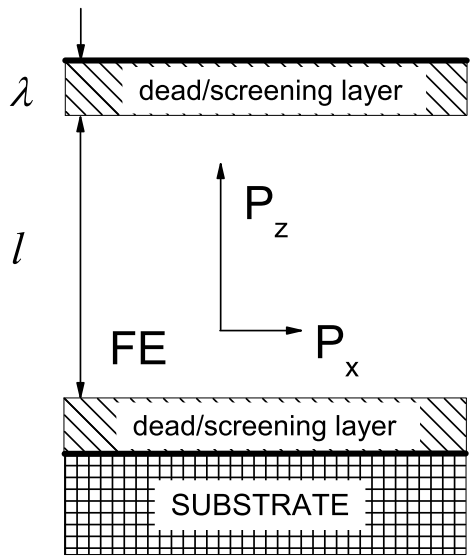}
\end{center}
\caption{Schematic of the (perovskite) ferroelectric film with thickness $l$%
\ and metal electrodes (with screening length $\protect\lambda $) on a
misfit substrate. The misfit makes the film a uniaxial ferroelectric with a
spontaneous polarization along $z-$axis.}
\label{fig:FEfilm}
\end{figure}

The main conclusions of this paper are made by analyzing the formula for
free energy of the total system as a function of the amplitude $a$ of the
ferroelectric "polarization wave" presenting the sinusoidal domain structure
and the homogeneous part of the ferroelectric polarization, $p$. For the
electrode and the film parameters of a system like BTO/SRO/STO, the
ferroelectric polarization that is perpendicular to the film plane,
schematic of which is shown in Fig.\ref{fig:FEfilm}, has the form
\begin{equation}
P_{z}\left( x,y,z\right) =p+a\cos \boldsymbol{kr}\cos qz,  \label{1}
\end{equation}%
where the orientation of $\boldsymbol{k}$ in the $x,y$ plane is not fixed, $%
2\pi /k$ is the period of the sinusoidal domain structure, $q=\pi /l$, and $%
l $ is the film thickness. To find the desired free energy, $F\left(
a,p\right) $, one has to find the elastic strains and the non-ferroelectric
polarization $\boldsymbol{P}_{\perp }=\left( P_{x},P_{y}\right) $ as
functions of $a$ and $p$ to present the total free energy as a function of $%
a $ and $p$ only. The total free energy contains contributions of the
ferroelectric film, of the substrate and of the electrode. In principle, it
should also contain a contribution of the voltage source but we consider
here a \emph{short-circuited} system and are not concerned with the latter
contribution.

When calculating elastic strains in the ferroelectric which accompany the
inhomogeneous polarization forming the sinusoidal domain structure, we
follow the same philosophy as in our previous work.\cite{BLKhach}. In
principle, when calculating these strains the inhomogeneous strains in the
substrate should be taken into account. However, it is well known that they
propagate into the substrate for about the same distances as the scale of
inhomogeneity in the film ($x,y$) plane. In our case, these inhomogeneities
are due to the domain structure, i.e. this scale is the period of the domain
structure. Then, it is convenient to consider relatively thick films since
the period of the domain structure is relatively small, specifically, it is
much less than the film thickness \cite{ChT82,BLreview}, Fig.~\ref%
{fig:FEonSub}. The contribution of the substrate is its elastic energy
which, as we have mentioned above, is concentrated within a volume which is
\emph{much smaller} than the film volume, as defined by a small factor $%
q/k=\pi /kl\ll 1$. Another convenience of the thick film limit is that it is
possible to disregard the boundary conditions for the inhomogeneous parts of
elastic strains and stresses at the surfaces of the ferroelectric. Indeed,
if we obtain a solution, which does not satisfy the boundary conditions, we
can find corrections to such a solution in a way that is conventionally used
in the elasticity theory. First, we apply the external forces to the
surfaces, which are necessary to meet the boundary conditions with the
strains corresponding to our solution making this solution correct. Second,
we apply forces opposite to the previous ones and find the strains produced
by the new forces. These strains provide the correction to the original
solution we were looking for. Once more, it is sufficient to observe that in
our case the external forces have the period of the domain structure to
understand that the elastic energy associated with the corrections necessary
to satisfy the boundary conditions can be neglected quite similarly to the
elastic energy of the substrate.
\begin{figure}[tbp]
\begin{center}
\includegraphics[angle=0,width=0.60\textwidth]{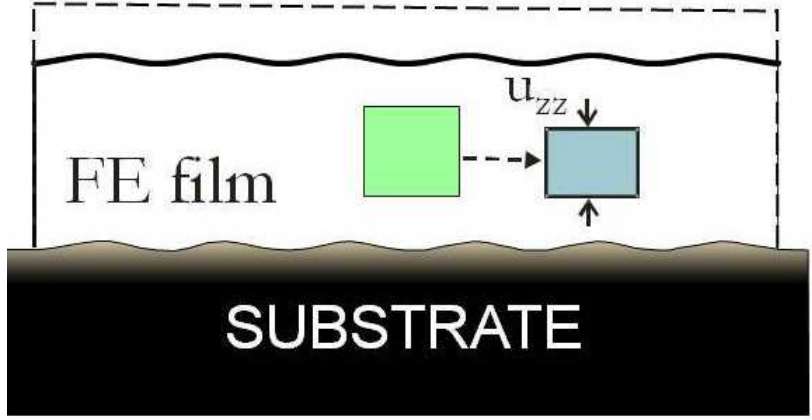}
\end{center}
\caption{(color online) Schematic of the ferroelectric film on the misfit
substrate at the onset of sinusoidal polarization wave. The elastic coupling
to the substrate allows inhomogeneous deformations but prohibits homogeneous
strains in plane of the film.}
\label{fig:FEonSub}
\end{figure}
Another convenience of thick film approximation is given by the possibility
to neglect those terms in the LGD free energy, which describe the
electrostriction but contain components of non-ferroelectric polarization.
According to Refs.\cite{ChT82,BLreview},
\begin{equation}
\boldsymbol{P}_{\perp }\left( x,y,z\right) =(\boldsymbol{k}/k)a_{\perp }\sin
\boldsymbol{kr}\sin qz,  \label{3}
\end{equation}%
where $a_{\perp }\approx aq/k$, $q/k=\pi /kl\ll 1.$The\ electrostriction
terms in the LGD free energy with non-ferroelectric components of
polarization may contain the ferroelectric component, like $%
P_{x}P_{z}u_{xz}, $ or may not contain them, like in the term $%
P_{x}P_{y}u_{xy}$. In both cases, they contribute to $a^{4}$ and $p^{2}a^{2}$
terms in the free energy depending on $a$ and $p$. In the first case, this
contribution is proportional to $\left( q/k\right) ^{2}$ and in the second
to $\left( q/k\right) ^{4}$. Since there are also the terms{\Large \ }$%
a^{4}, $ $p^{2}a^{2}$ that do not contain the small factor $q/k,$ the
contribution of these terms can be neglected.

Taking this into account, we write down the LGD free energy in the form:%
\begin{equation}
F\left( \boldsymbol{P},u_{ik}\right) =F_{1}\left( \boldsymbol{P}\right)
+F_{2}\left( u_{ik}\right) +F_{3}\left( \boldsymbol{P},u_{ik}\right) ,
\label{6}
\end{equation}%
where%
\begin{equation}
F_{1}\left( \boldsymbol{P}\right) =\frac{A}{2}P_{z}^{2}+\frac{B}{4}P_{z}^{4}+%
\frac{1}{2}G\left( \nabla _{\bot }P_{z}\right) ^{2}+\frac{1}{2}\kappa
P_{bz}^{2}+\frac{A_{\perp }}{2}P_{\bot }^{2},  \label{7}
\end{equation}%
\begin{equation}
F_{2}\left( u_{ik}\right) =\frac{1}{2}\lambda _{1}\left(
u_{xx}^{2}+u_{yy}^{2}+u_{zz}^{2}\right) +\lambda _{2}\left(
u_{xx}u_{yy}+u_{xx}u_{zz}+u_{zz}u_{yy}\right) +2\mu \left(
u_{xy}^{2}+u_{zy}^{2}+u_{xz}^{2}\right) ,  \label{8}
\end{equation}%
\begin{equation}
F_{3}\left( \boldsymbol{P},u_{ik}\right) =q_{11}u_{zz}P_{z}^{2}+q_{12}\left(
u_{xx}+u_{yy}\right) P_{z}^{2}.  \label{9}
\end{equation}%
Here, $q_{11(12)}$ are the standard piezo-electric coefficients that should
not be confused with the parameter $q$ defining the transversal profile of
the polarization wave (\ref{1}). In Eq.(\ref{7}), $A=\gamma \left(
T-T_{c}\right) ,$ $B,G=\mathrm{const,}$ $\overrightarrow{\boldsymbol{\nabla }%
}_{\bot }\mathrm{=(\partial }_{x},\partial _{y})$ the gradient in the plane
of the film, $P_{bz}$ is the non-ferroelectric (`base') part of the
polarization perpendicular to the electrodes \cite{Tagantsev}, $A_{\bot }>0.$
Following Refs \cite{ChT82,BLreview}, we have neglected a term with the
gradient in $z-$direction since it is much smaller than the one in plane of
the film, $\partial _{z}\ll \overrightarrow{\boldsymbol{\nabla }}_{\bot }$.
It is worth mentioning that we have not included the energy of the electric
field.into the LGD free energy. The reason is that we shall use it to write
down the constituent equations only. We shall eliminate $u_{ik\text{ }}$, $%
\boldsymbol{P}_{\bot },$ $P_{bz}$ as well the electric field components from
the system of constituent electrostatics equations to obtain two coupled
equations of state for $a$ and $p$. We shall obtain $F\left( p,a\right) $
from the resulting equations. This is possible because of the thick films
approximation. The most straightforward method to obtain $F\left( p,a\right)
$\ would be to substitute Eq.(\ref{1}) into Eq.(\ref{6}) supplemented by the
electric field energy and to integrate over the film volume. In general, the
result would not be the same as the one obtained from the constituent
equations because of approximate character of Eq.(\ref{1}). However, for $%
q=\pi /l$\ the two results coincide and that makes it possible to use a more
convenient method of the constituent equations.

\section{Constituent equations\label{sec:constit}}

For the polarization components one has:%
\begin{equation}
AP_{z}+BP_{z}^{3}-G\triangledown _{\bot
}^{2}P_{z}+2q_{11}P_{z}u_{zz}+2q_{12}P_{z}\left( u_{xx}+u_{yy}\right) =E_{z},
\label{10}
\end{equation}%
\begin{eqnarray}
P_{bz} &=&\kappa E_{z},  \label{eq:Pbx} \\
\boldsymbol{P}_{\perp } &=&A_{\bot }\boldsymbol{E}_{\bot }.  \label{eq:Pperp}
\end{eqnarray}%
Before writing down the equations for the strain, we shall eliminate the
electric field from the above three equations. Assuming Eq.(\ref{1}) for $%
P_{z}$, Eq.(\ref{3}) for $\boldsymbol{P}_{\perp }$ and putting \cite%
{BLreview}:%
\begin{equation}
E_{0z}=E_{0}+E_{z}^{k}\cos \boldsymbol{kr}\cos qz,\text{ \ }\boldsymbol{E}%
_{\bot }=(\boldsymbol{k}/k)E_{\perp }^{k}\sin \boldsymbol{kr}\sin qz
\label{12}
\end{equation}%
we can replace Eqs.(\ref{10}),(\ref{eq:Pperp}) with
\begin{equation}
Ap+\left[ BP_{z}^{3}+2q_{11}P_{z}u_{zz}+2q_{12}P_{z}u_{\perp \perp }\right]
_{\hom }=E_{0},  \label{13}
\end{equation}%
\begin{equation}
\left( A+Gk^{2}\right) a+\left[
BP_{z}^{3}+2q_{11}P_{z}u_{zz}+2q_{12}P_{z}u_{\perp \perp }\right] _{\mathrm{%
cc}}=E_{z}^{k},  \label{14}
\end{equation}%
\begin{equation}
A_{\perp }a_{\perp }=E_{\perp }^{k},  \label{15}
\end{equation}%
where $u_{\perp \perp }=u_{xx}$ $+u_{yy}$, $\left[ \ldots \right] _{\hom }$
and $\left[ \ldots \right] _{\mathrm{cc}}$ denote the homogeneous part ($%
k=0) $ and the part proportional to $\cos \boldsymbol{kr}\cos qz$ of the
expression in the brackets, correspondingly. Of course, as a result of this
replacement, a part of the l.h.s. of Eq.(\ref{10}) is lost but it
corresponds to the higher harmonics of the sinusoidal distribution of the
polarization and these harmonics can be neglected close to the transition
\cite{ChT82,BLreview}.

The homogeneous part of the electric field $E_{z0}$ can be calculated as,
e.g., in Ref.\cite{BLreview} yielding for the short-circuited case
\begin{equation}
E_{0z}=-\frac{4\pi d}{\varepsilon _{b}d+\epsilon _{e}l}p,  \label{22}
\end{equation}%
where $d$ is the thickness of the dead layer and $\epsilon _{e}$ its
dielectric constant. Recall that real electrodes have finite albeit small
(Thomas-Fermi) screening length $\lambda ,$ which is completely analogous
\cite{BLreview} to a presence of the `dead' non-ferroelectric layers at the
interface with thickness $d/2=\lambda .$ Using Eq. (\ref{22}), the equation (%
\ref{13}) gets the form:%
\begin{equation}
A_{1}p+\left[ BP_{z}^{3}+2q_{11}P_{z}u_{zz}+2q_{12}P_{z}\left(
u_{xx}+u_{yy}\right) \right] _{\hom }=0.  \label{23}
\end{equation}%
where%
\begin{equation}
A_{1}=A+\frac{4\pi d}{\varepsilon _{b}d+\epsilon _{e}l}\approx A+\frac{4\pi d%
}{\epsilon _{e}l},  \label{47}
\end{equation}%
since usually the dead layer is very thin, $\varepsilon _{b}d\ll \epsilon
_{e}l.$ To transform Eqs.~(\ref{14}) we use the electrostatics equation,
\begin{equation}
\mathrm{div}\boldsymbol{D}=0\mathrm{,}  \label{16}
\end{equation}%
where $\mathbf{D}$ is the dielectric displacement, firstly for the
ferroelectric material, taking into account that $\boldsymbol{D}=\left(
\varepsilon _{\perp }\boldsymbol{E}_{\bot },\quad \varepsilon _{b}E_{z}+4\pi
P_{z}\right) $, where $\varepsilon _{\perp }=1+4\pi /A_{\perp }$, and $%
\varepsilon _{b}=1+4\pi /\kappa $ is the `\emph{base}' non-critical
dielectric constant\cite{Tagantsev,BLreview}, and together with the equation
$\mathrm{curl}\boldsymbol{E}=0,$ we find that

\begin{equation}
E_{z}^{k}=-\frac{4\pi q^{2}}{\varepsilon _{\perp }k^{2}}a.  \label{19}
\end{equation}%
Substituting (\ref{19}) into the equation for the amplitude of the
`polarization wave' $a,$ Eq.(\ref{14}), we rewrite the latter as:
\begin{equation}
\left( A+Gk^{2}+\frac{4\pi q^{2}}{\varepsilon _{\perp }k^{2}}\right) a+\left[
BP_{z}^{3}+2q_{11}P_{z}u_{zz}+2q_{12}P_{z}u_{\perp \perp }\right] _{\mathrm{%
cc}}=0.
\end{equation}%
The nontrivial solution of the above equation first appears when the
coefficient in the first term in round brackets before the amplitude $a$ in
the above equation first crosses zero, i.e. when $A=\left[ -Gk^{2}-4\pi
q^{2}/\left( \varepsilon _{\perp }k^{2}\right) \right] _{\max }$. That takes
place at some $A<0$, so upon lowering temperature at constant thickness the
transition into domain state occurs somewhat below the bulk critical
temperature $T_{c}$, in other words. The transition for varying thickness of
the film at some constant temperature $T<T_c$ takes place when the thickness
exceeds some critical value. Therefore, the first nontrivial solution
appears for the `polarization wave'\cite{ChT82,BLreview} with the wave
number $k$ that minimizes the sum $Gk^{2}+4\pi q^{2}/\left( \varepsilon
_{\perp }k^{2}\right) $, so that (recall that $q=\pi /l$)
\begin{equation}
\frac{4\pi q^{2}}{\varepsilon _{\perp }k^{2}}=Gk^{2},\quad k=\left( \frac{%
4\pi q^{2}}{\varepsilon _{\perp }G}\right) ^{1/4}=\left( \frac{4\pi ^{3}}{%
\varepsilon _{\perp }Gl^{2}}\right) ^{1/4}.  \label{eq:k}
\end{equation}%
We can now rewrite Eq.(\ref{14}) as the homogeneous one:%
\begin{equation}
A_{2}a+\left[ BP_{z}^{3}+2q_{11}P_{z}u_{zz}+2q_{12}P_{z}u_{\perp \perp }%
\right] _{\mathrm{cc}}=0,  \label{21}
\end{equation}%
where
\begin{equation}
A_{2}=A+2Gk^{2}.  \label{48}
\end{equation}

It is seen from Eq.~(\ref{9}) that the only source of elastic stresses and
strains is $P_{z}^{2}\left( x,y,z\right) $ in our approximation. Since%
\begin{equation}
P_{z}^{2}=p^{2}+2pa\cos \boldsymbol{kr}\cos qz+\frac{a^{2}}{4}\left( 1+\cos
2qz+\cos 2\boldsymbol{kr}+\cos 2\boldsymbol{kr}\cos 2qz\right) ,  \label{28a}
\end{equation}%
we should expect that%
\begin{equation}
u_{zz}=u_{zz}^{\left( 0\right) }+u_{zz}^{\left( 1\right) }\cos
2qz+u_{zz}^{(2)}\cos \boldsymbol{kr}\cos qz+u_{zz}^{(3)}\cos 2\boldsymbol{kr}%
+u_{zz}^{(4)}\cos 2\boldsymbol{kr}\cos 2qz,  \label{29}
\end{equation}%
while for $u_{xx}$, $u_{yy},$ and for $u_{\perp \perp }$ we shall have
similar formulas with the homogeneous part (first term in the above
expression) absent because of the substrate. The superscripts $\left(
0\right) -\left( 4\right) $\ denote contributions with different types of
the coordinate dependencies as defined by Eq.(\ref{29}). Below, we use the
same superscripts for both the coefficients and the functions.

Substituting Eqs.(\ref{28a}),(\ref{29}) and analogous equations for $u_{xx}$
and $u_{yy}$ into Eqs.(\ref{23}),(\ref{21}), we find:
\begin{equation}
A_{1}p+\left[ BP_{z}^{3}\right] _{\hom }+2q_{11}\left( pu_{zz}^{\left(
0\right) }+\frac{au_{zz}^{(2)}}{4}\right) +q_{12}\frac{a}{2}u_{\perp \perp
}^{(2)}=0,  \label{30}
\end{equation}

\begin{equation}
A_{2}a+\left[ BP_{z}^{3}\right] _{\mathrm{cc}}+2q_{11}\left[
pu_{zz}^{(2)}+a\left( u_{zz}^{\left( 0\right) }+\frac{u_{zz}^{\left(
1\right) }+u_{zz}^{\left( 3\right) }}{2}+\frac{u_{zz}^{\left( 4\right) }}{4}%
\right) \right] \text{ }+2q_{12}\left[ pu_{\perp \perp }^{(2)}+a\left( \frac{%
u_{\perp \perp }^{(1)}+u_{\perp \perp }^{(3)}}{2}+\frac{u_{\perp \perp
}^{(4)}}{4}\right) \right] =0.  \label{31}
\end{equation}%
We shall calculate the values $u_{ik}^{\left( j\right) }$ in the next
Section by solving the elastic problem explicitly.

Importantly, the above equation of state (\ref{30})\ suggests that the film
would tend to transform into a single domain (SD)\ state with $p\neq 0$ and $%
a=0$ at temperature $T_{c}^{SD}$ such that $A_{1}=0$ or, in other words,
\begin{equation}
A\left( T_{c}^{SD}\right) =-4\pi d/\left( \epsilon _{e}l\right) .
\label{eq:TcSD}
\end{equation}%
\ The second equation of state (\ref{31}) yields a transition into a domain
state ($p=0$ and $a\neq 0)$ at the temperature $T_{d}$ such that $A_{2}=0.$
or%
\begin{equation}
A(T_{d})=-2Gk^{2}=-4\left( \frac{\pi ^{3}G}{\varepsilon _{\perp }}\right)
^{1/2}\frac{1}{l}.  \label{eq:Td}
\end{equation}%
Recall that in the present case, corresponding to BaTiO$_{3}$/SrRuO$_{3}$%
/SrTiO$_{3}$\cite{BLapl06},%
\begin{equation}
\frac{4\pi d}{\epsilon _{e}l}>2Gk^{2}\sim \frac{4\pi d_{at}}{\varepsilon
_{\perp }^{1/2}l},  \label{23a}
\end{equation}%
where $d_{at}=\sqrt{\pi G}\approx 1\mathring{A}$ is the small `atomic'
length scale ($G=0.3\mathring{A}^{2}$ for BaTiO$_{3}$\cite{BLapl06,BLreview}%
). The above relation means that the paraphase gives way to the \emph{domain
phase}, with $a\neq 0,$ thus preventing it from reaching the temperature $%
T_{d}$ where it could have transformed into a single domain state.
Obviously, same is true of the phase transformations in the film as a
function of \emph{thickness} at constant temperature. There, one can
introduce the critical thickness for domains, $l_{d},$ where \
\begin{equation}
A=-2Gk^{2}=-4\left( \frac{\pi ^{3}G}{\varepsilon _{\perp }}\right) ^{1/2}%
\frac{1}{l_{d}},  \label{eq:ld}
\end{equation}%
and the `critical thickness for the single domain state' $l_{c}^{SD}$, such
that
\begin{equation}
A=-4\pi d/\left( \epsilon _{e}l_{c}^{SD}\right) .  \label{eq:lcSD}
\end{equation}%
These introduced critical thicknesses and temperatures are discussed in
detail below in Sec.\ref{sec:mdloss}.

\section{Elastic problem{\label{sec:elproblem}}}

Using Eqs.~(\ref{8}),(\ref{9}), we obtain for the diagonal components of the
elastic stress tensor:%
\begin{equation}
\sigma _{xx}=\lambda _{1}u_{xx}+\lambda _{2}\left( u_{yy}+u_{zz}\right)
+q_{12}P_{z}^{2},  \label{24}
\end{equation}%
\begin{equation}
\sigma _{yy}=\lambda _{1}u_{yy}+\lambda _{2}\left( u_{xx}+u_{zz}\right)
+q_{12}P_{z}^{2},  \label{25}
\end{equation}%
\begin{equation}
\sigma _{zz}=\lambda _{1}u_{zz}+\lambda _{2}\left( u_{xx}+u_{yy}\right)
+q_{11}P_{z}^{2},  \label{26}
\end{equation}%
and formulas of the type%
\begin{equation}
\sigma _{xy}=2\mu u_{xy},  \label{27}
\end{equation}%
for the off-diagonal components.

We have already mentioned that the only $u_{ik}$ component which has a
non-zero homogeneous part is $u_{zz}.$ This part is easily found from the
condition at the free surface: $\sigma _{zz}=0$ at $z=l/2$. From Eq.(\ref{26}%
), one finds:%
\begin{equation}
u_{zz}^{\left( 0\right) }=-\frac{q_{11}}{\lambda _{1}}\left[ P_{z}^{2}\right]
_{\hom }=-\frac{q_{11}}{\lambda _{1}}\left( p^{2}+\frac{a^{2}}{4}\right) .
\label{32}
\end{equation}%
For the parts depending on $z$ \emph{only}, the equations of elastic
equilibrium take the form:

\begin{equation}
\partial \sigma _{iz}^{(1)}/\partial z=0,  \label{33}
\end{equation}%
i.e. $\sigma _{iz}^{(1)}=\mathrm{const}=0$ since it should vanish at the
free surface ($z=l/2)$. Therefore, Eqs.~(\ref{26}),(\ref{28a}) yield%
\begin{equation}
u_{zz}^{\left( 1\right) }=-q_{11}a^{2}/\left( 4\lambda _{1}\right) ,
\label{35}
\end{equation}%
and
\begin{equation}
u_{xx}^{\left( 1\right) }=u_{yy}^{\left( 1\right) }=0,
\label{eq:uxx1_uyy1=0}
\end{equation}%
because of the Saint-Venant's elastic compatibility conditions for $z-$only
dependent strains.

When solving the rest of the elastic problem, we shall use\ the small
parameter $q/k\ll 1$. This allows us to neglect the derivatives with respect
to $z$: formally, $\partial /\partial z\ll \partial /\partial x,\partial
/\partial y.$ As a result, the equations of the elastic equilibrium acquire
the form:%
\begin{equation}
\frac{\partial \sigma _{xx}^{\left( 2-4\right) }}{\partial x}+\frac{\partial
\sigma _{xy}^{\left( 2-4\right) }}{\partial y}=\frac{\partial \sigma
_{yz}^{\left( 2-4\right) }}{\partial y}+\frac{\partial \sigma _{zx}^{\left(
2-4\right) }}{\partial x}=\frac{\partial \sigma _{yy}^{\left( 2-4\right) }}{%
\partial y}+\frac{\partial \sigma _{yx}^{\left( 2-4\right) }}{\partial x}=0,
\label{36}
\end{equation}%
where the superscripts $\left( 2-4\right) $ denote the part of stresses that
is due to the three last terms in Eq.(\ref{28a}), which we denote as $%
P_{z}^{2\left( 2-4\right) }$. Explicitly,%
\begin{equation}
\lambda _{1}\frac{\partial ^{2}u_{x}^{\left( 2-4\right) }}{\partial x^{2}}%
+\lambda _{2}\frac{\partial ^{2}u_{y}^{\left( 2-4\right) }}{\partial
y\partial x}+\mu \frac{\partial ^{2}u_{x}^{\left( 2-4\right) }}{\partial
y^{2}}+\mu \frac{\partial ^{2}u_{y}^{\left( 2-4\right) }}{\partial y\partial
x}+q_{12}\frac{\partial P_{z}^{2\left( 2-4\right) }}{\partial x}=0,
\label{37}
\end{equation}%
\begin{equation}
\mu \frac{\partial ^{2}u_{z}^{\left( 2-4\right) }}{\partial y^{2}}+\mu \frac{%
\partial ^{2}u_{z}^{\left( 2-4\right) }}{\partial x^{2}}=0,  \label{38}
\end{equation}%
\begin{equation}
\lambda _{1}\frac{\partial ^{2}u_{y}^{\left( 2-4\right) }}{\partial y^{2}}%
+\lambda _{2}\frac{\partial ^{2}u_{x}^{\left( 2-4\right) }}{\partial
y\partial x}+\mu \frac{\partial u_{x}^{\left( 2-4\right) }}{\partial
x\partial y}+\mu \frac{\partial ^{2}u_{y}^{\left( 2-4\right) }}{\partial
x^{2}}+q_{12}\frac{\partial P_{z}^{2\left( 2-4\right) }}{\partial y}=0.
\label{39}
\end{equation}%
Analogously to the isotropic case \cite{BLKhach}, we shall put the
conditions $u_{z}^{\left( 2-4\right) }=0$ that satisfy Eq.(\ref{38}) but
not, of course, the boundary conditions. The latter is not important in our
approximation, as we argued above. Therefore, we conclude that
\begin{equation}
u_{zz}^{(2)}=u_{zz}^{(3)}=u_{zz}^{(4)}=0,  \label{41}
\end{equation}%
and we are left with only two equations to solve. It is convenient to solve
them separately for $\left( 2\right) $ and $\left( 3,4\right) $ parts, since
they correspond to different spatial harmonics.

Simplifying the remaining equations (\ref{37}),(\ref{39}), we obtain:
\begin{equation}
\lambda _{1}\frac{\partial ^{2}u_{x}^{\left( 2-4\right) }}{\partial x^{2}}%
+(\lambda _{2}+\mu )\frac{\partial ^{2}u_{y}^{\left( 2-4\right) }}{\partial
y\partial x}+\mu \frac{\partial ^{2}u_{x}^{\left( 2-4\right) }}{\partial
y^{2}}+q_{12}\frac{\partial P_{z}^{2\left( 2-4\right) }}{\partial x}=0,
\end{equation}%
\begin{equation}
\lambda _{1}\frac{\partial ^{2}u_{y}^{\left( 2-4\right) }}{\partial y^{2}}%
+(\lambda _{2}+\mu )\frac{\partial ^{2}u_{x}^{\left( 2-4\right) }}{\partial
y\partial x}+\mu \frac{\partial ^{2}u_{y}^{\left( 2-4\right) }}{\partial
x^{2}}+q_{12}\frac{\partial P_{z}^{2\left( 2-4\right) }}{\partial y}=0.
\end{equation}%
For the terms $u^{(2)},$ we have $P_{z}^{2(2)}\propto \cos \boldsymbol{kr},$
$\partial _{x(y)}P_{z}^{2}\propto -k_{x(y)}\sin \boldsymbol{kr},$ meaning
that $u_{x(y)}\propto \sin \boldsymbol{kr},$ $\ \partial ^{2}u_{i}/\partial
x^{2}=-k_{x}^{2}u_{i}$, etc. Then,%
\begin{equation}
\lambda _{1}k_{x}^{2}u_{x}^{\left( 2\right) }+(\lambda _{2}+\mu
)k_{x}k_{y}u_{y}^{\left( 2\right) }+\mu k_{y}^{2}u_{x}^{\left( 2\right)
}+q_{12}k_{x}2pa=0,
\end{equation}%
\begin{equation}
\lambda _{1}k_{y}^{2}u_{y}^{\left( 2\right) }+\left( \lambda _{2}+\mu
\right) k_{x}k_{y}u_{x}^{\left( 2\right) }+\mu k_{x}^{2}u_{y}^{\left(
2\right) }+q_{12}k_{y}2pa=0.
\end{equation}%
The terms $u^{(3),(4)}$\ correspond to higher spatial harmonics in Eq.~(\ref%
{28a}), but they should be taken into account since they contribute to the
terms with the main harmonic in the constituent equation (\ref{31}). Since$\
$for\ this part $P_{z}^{2}\propto \cos 2kr${\Large ,} we obtain a slightly
different set of equations:

\begin{eqnarray}
\lambda _{1}k_{x}^{2}u_{x}^{\left( 3,4\right) }+(\lambda _{2}+\mu
)k_{x}k_{y}u_{y}^{\left( 3,4\right) }+\mu k_{y}^{2}u_{x}^{\left( 3,4\right)
}+q_{12}k_{x}\frac{a^{2}}{8} &=&0, \\
\lambda _{1}k_{y}^{2}u_{y}^{\left( 3,4\right) }+\left( \lambda _{2}+\mu
\right) k_{x}k_{y}u_{x}^{\left( 3,4\right) }+\mu k_{x}^{2}u_{y}^{\left(
3,4\right) }+q_{12}k_{y}\frac{a^{2}}{8} &=&0.
\end{eqnarray}%
Note that for the constituent equations we need the combinations:
\begin{eqnarray*}
u_{\perp \perp }^{(2)} &=&u_{xx}^{\left( 2\right) }+u_{yy}^{\left( 2\right)
}=k_{x}u_{x}^{(2)}+k_{y}u_{y}^{(2)}, \\
u_{\perp \perp }^{(3,4)} &=&u_{xx}^{(3,4)}+u_{yy}^{\left( 3,4\right)
}=2k_{x}u_{x}^{(3,4)}+2k_{y}u_{y}^{(3,4)}.
\end{eqnarray*}%
We find:%
\begin{equation}
u_{\perp \perp }^{(2)}=-2q_{12}apf\left( \theta \right) ,  \label{42}
\end{equation}%
where $\theta $ is defined by $k_{x}=k\cos \theta $, \ $k_{y}=k\sin \theta $
and the function $f\left( \theta \right) $ is

\begin{equation}
f\left( \theta \right) =2\frac{\left( \lambda _{1}-\lambda _{2}\right) \sin
^{2}2\theta +2\mu \cos ^{2}2\theta }{\left( \lambda _{1}+\lambda _{2}+2\mu
\right) \left( \lambda _{1}-\lambda _{2}\right) \sin ^{2}2\theta +4\lambda
_{1}\mu \cos ^{2}2\theta }.  \label{43}
\end{equation}%
For terms, corresponding to $\cos 2\boldsymbol{k}\boldsymbol{r}$ and $\cos
2qz,$ we obtain%
\begin{equation}
u_{\perp \perp }^{(3)}=u_{\perp \perp }^{(4)}=-q_{12}\frac{a^{2}}{4}f\left(
\theta \right) .  \label{44}
\end{equation}

Using the results of solution of the elastic problem, Eqs.(\ref{32}),(\ref%
{41}),(\ref{42}),(\ref{44}), we can write Eqs.(\ref{30}),(\ref{31}) as%
\begin{equation}
A_{1}p+\left( B-\frac{2q_{11}^{2}}{\lambda _{1}}\right) p^{3}+pa^{2}\frac{3}{%
4}\left[ B-\frac{4}{3}\left( \frac{q_{11}^{2}}{2\lambda _{1}}%
+q_{12}^{2}f\left( \theta \right) \right) \right] =0  \label{45}
\end{equation}%
\begin{equation}
A_{2}a+a^{3}\frac{9}{16}\left( B-\frac{4}{3}\left[ \frac{q_{11}^{2}}{\lambda
_{1}}+\frac{q_{12}^{2}}{2}f\left( \theta \right) \right] \right)
+ap^{2}3\left( B-\frac{4}{3}\left[ \frac{q_{11}^{2}}{2\lambda _{1}}%
+q_{12}^{2}f\left( \theta \right) \right] \right) =0.  \label{46}
\end{equation}%
From these two constituent equations corresponding to an extremum of the
free energy, one can easily reconstruct the free energy:
\begin{equation}
V^{-1}\tilde{F}(p,a)=\frac{A_{1}}{2}p^{2}+\frac{A_{2}}{8}a^{2}+\frac{%
\widetilde{B}}{4}p^{4}+\frac{3B_{1}\left( \theta \right) }{8}a^{2}p^{2}+%
\frac{9B_{2}\left( \theta \right) }{256}a^{4},  \label{eq:Fpa}
\end{equation}%
where%
\begin{equation}
\widetilde{B}=B-\frac{2q_{11}^{2}}{\lambda _{1}},\quad B_{1}\left( \theta
\right) =\widetilde{B}+\frac{4}{3}\left[ \frac{q_{11}^{2}}{\lambda _{1}}%
-q_{12}^{2}f\left( \theta \right) \right] ,\quad B_{2}\left( \theta \right) =%
\widetilde{B}+\frac{2}{3}\left[ \frac{q_{11}^{2}}{\lambda _{1}}%
-q_{12}^{2}f\left( \theta \right) \right] ,  \label{49}
\end{equation}%
are the Landau coefficients before the quartic terms renormalized by the
strain. The form of the free energy is the same as in the isotropic case
\cite{ChT82,BLreview,BLKhach}, but, importantly, the coefficients $B_{1}$
and $B_{2}$ depend on the \emph{orientation} of the `polarization wave'
given by the angle $\theta $.

\section{Orientation of the domain structure\label{sec:orient}}

Consider the domain structure formed close to the paraelectric-ferroelectric
transition. Although stability of the paraelectric phase is lost with
respect to the polarization waves with the value of the $\boldsymbol{k}$
vector given by Eq.(\ref{eq:k}) and arbitrary orientation in the $x-y$
plane, i.e. for any $\theta ,$ the energy of sinusoidal domain structure
depends on $\theta $ and one has to find the ones corresponding to the
equilibrium domain structure(s). Since we consider here one `polarization
wave' only, we study the competition between the stripe-type structures. In
Sec.\ref{sec:checker} we shall show that the square (checkerboard) domain
structure is unstable in perovskite crystals that we study here. The
checkerboard structure could, in principle, be stable or metastable under
some conditions on the material constants, but we are not aware of any
experimental example of this type, so it will be premature to study such a
hypothetical case.

Recall that we discuss the ferroelectric phase transition in a sample\ with
short-circuited electrodes. Then, $p=0$ in the ferroelectric phase at least
not far from the phase transition, and the phase transition into the
inhomogeneous domain phase occurs at $A_{2}=0$. The free energy is:%
\begin{equation}
V^{-1}\tilde{F}(p,a)=\frac{A_{2}}{8}a^{2}+\frac{9B_{2}\left( \theta \right)
}{256}a^{4}.  \label{52}
\end{equation}%
At a fixed $\theta ,$ the minimum of this free energy is realized for
\begin{equation}
a^{2}=-\frac{16A_{2}}{9B_{2}\left( \theta \right) },  \label{53}
\end{equation}%
with the corresponding free energy
\begin{equation}
V^{-1}\tilde{F}_{\min }(p,a)=-\frac{A_{2}^{2}}{9B_{2}\left( \theta \right) }.
\label{54}
\end{equation}%
We see that the equilibrium domain structure is realized for the angles $%
\theta $ which minimize the function $B_{2}\left( \theta \right) $ or,
according to Eq.(\ref{49}), \emph{maximize} the function $f\left( \theta
\right) $. Let us find maxima of this function. It can be written in the
form:%
\begin{equation}
f\left( \theta \right) =\frac{2}{\lambda _{1}+\lambda _{2}+2\mu }\left( 1+%
\frac{r-c}{\tan ^{2}2\theta +c}\right) ,  \label{55}
\end{equation}%
where%
\begin{equation}
r=\frac{2\mu }{\lambda _{1}-\lambda _{2}},\text{ \ }c=\frac{4\lambda _{1}\mu
}{\left( \lambda _{1}+\lambda _{2}+2\mu \right) \left( \lambda _{1}-\lambda
_{2}\right) }.  \label{56}
\end{equation}%
One sees that for $r>c$ or
\begin{equation}
\lambda _{2}+2\mu >\lambda _{1},  \label{57}
\end{equation}%
the equilibrium domain structure corresponds to $\theta _{eq}=0,\quad \pi /2$
\ and%
\begin{equation*}
f\left( \theta \right) _{\max }=f(0)=1/\lambda _{1},
\end{equation*}%
while in the opposite case \ $\theta _{eq}=\pi /4,\quad 3\pi /4$ and
\begin{equation}
f\left( \theta \right) _{\max }=f\left( \pi /4\right) =\frac{2}{\lambda
_{1}+\lambda _{2}+2\mu }.
\end{equation}%
Throughout the present paper, we use the data for the material constants of
BaTiO$_{3}$ and PbTiO$_{3}$ from Refs \cite%
{Pertsev98,perkohl06,Li02,Sheng08,Hlinka} and of Pb(Zr$_{0.5}$Ti$_{0.5}$)O$%
_{3}$ from Ref.\cite{perkohl06}. We see that for BTO and PTO the condition
of Eq.(\ref{57}) is met and, therefore, the equilibrium $180^{\circ }$
domain structure consists of stripes parallel (perpendicular) to the cubic
axes, while the opposite inequality applies to PZT and the stripes make $%
45^{\circ }$ with the cubic axes there. For BTO, our conclusion coincides
with that of Dvorak and Janovec \cite{Dvorak} who defined the equilibrium
orientation of the $180^{\circ }$ domain walls in BTO far from the phase
transition. These authors were surprised by their conclusion about a very
weak orientational dependence of the domain wall energy given that the
experimental observations \cite{Fousek} showed a clearly preferable
orientation, the same as suggested by the theory.

It follows from our results that the weak orientational dependence of the
domain structure energy takes place in the sinusoidal regime too, and not
only for BTO, but for all three perovskites we have made the numerical
estimates for.
\begin{figure}[tbp]
\begin{center}
\includegraphics[angle=0,width=0.60\textwidth]{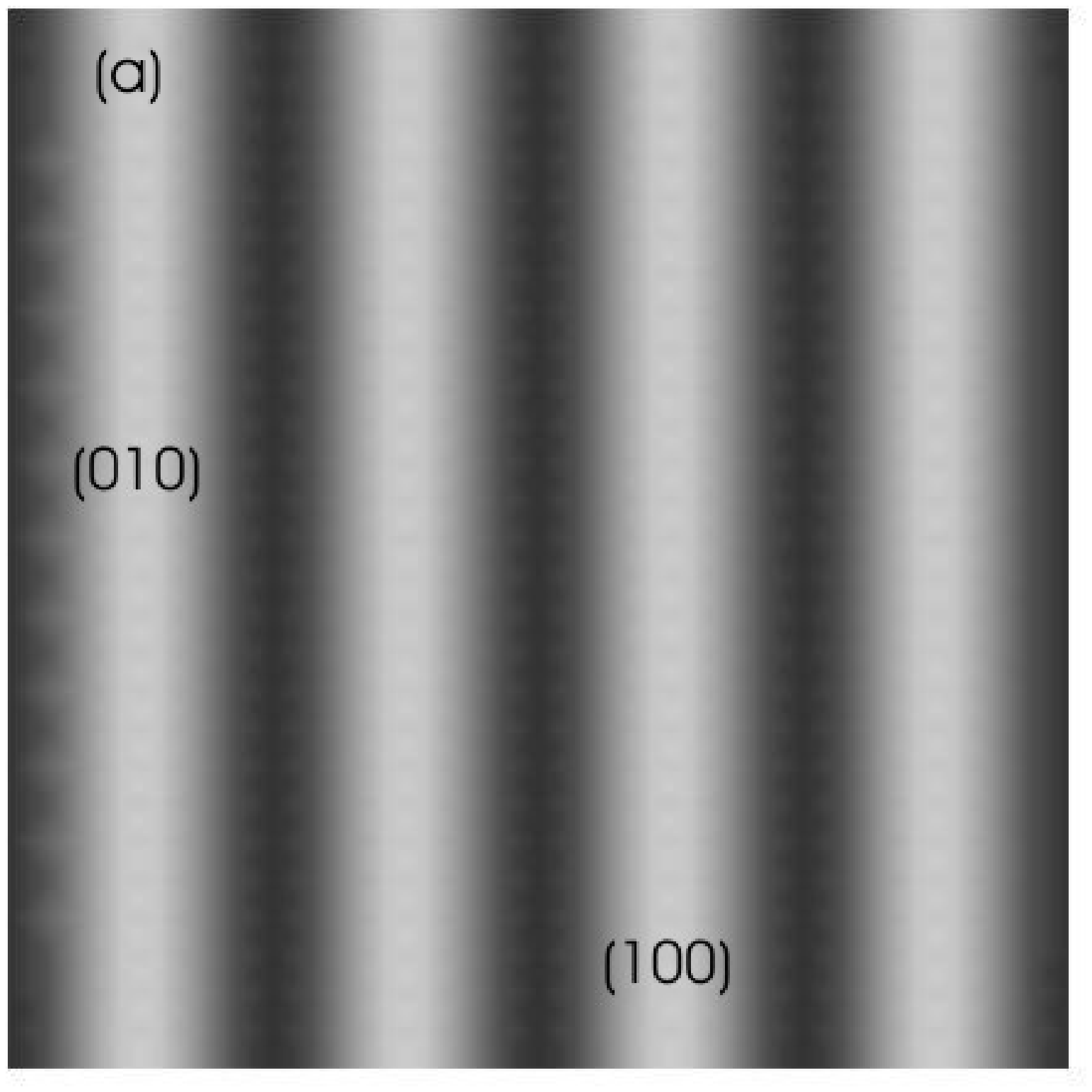} %
\includegraphics[angle=0,width=0.60\textwidth]{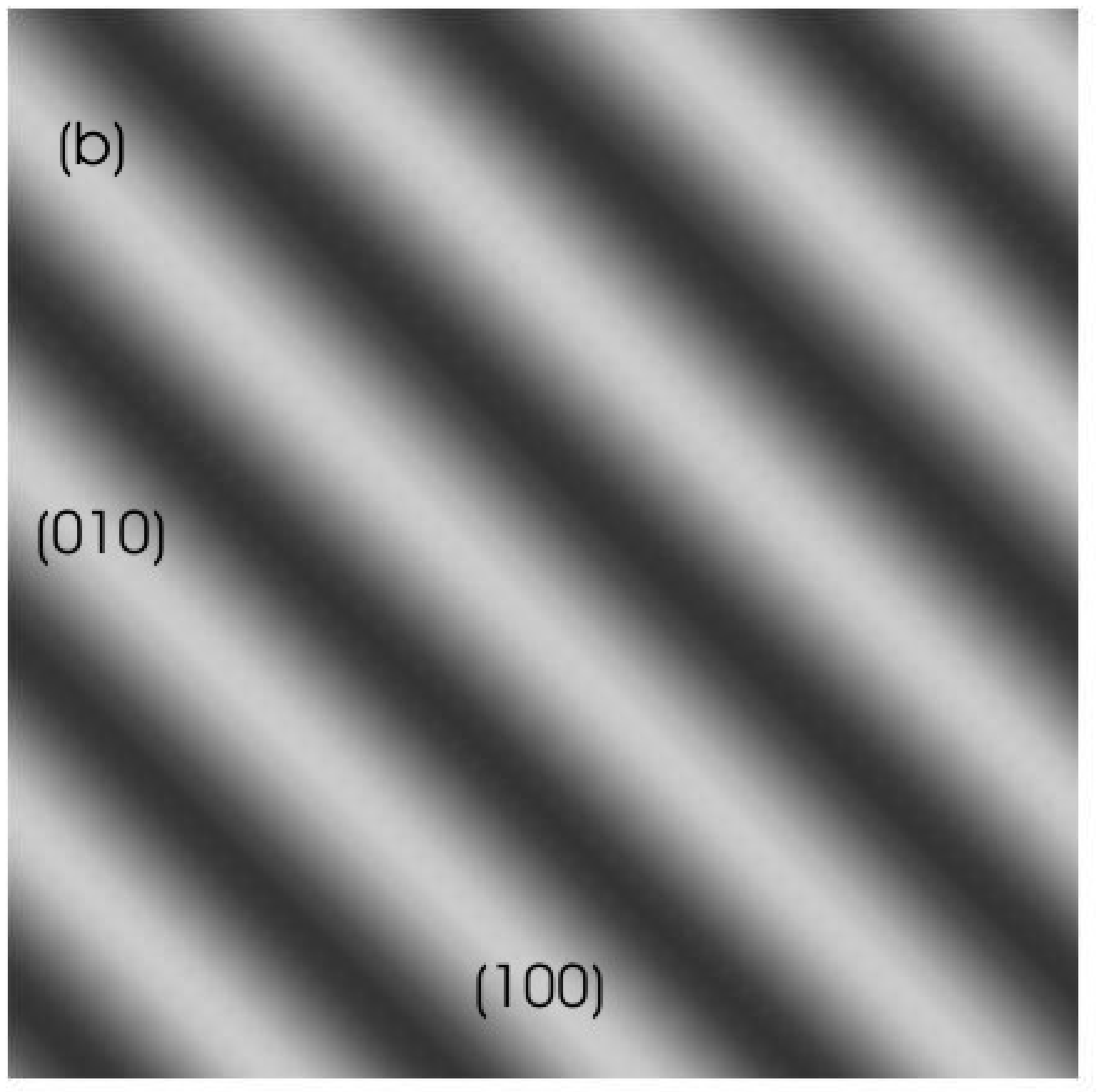}
\end{center}
\caption{Schematic of the sinusoidal domain structure in (a) BTO and PTO,
and (b) PZT. While the stripes are oriented along the crystal axes in case
(a), in case (b) the stripes are at 45 degrees with respect to cubic axes,
the difference being due to an opposite sign of elastic anizotropy in those
two cases. }
\label{fig:stripes}
\end{figure}
Indeed, from Eq.~(\ref{54}) one sees that the orientational dependence of
the domain structure energy comes from the function $B_{2}\left( \theta
\right) . $ The maximum difference of values Eq.(\ref{49}) for $B_{2}(\theta
)$ can be used to characterize the anizotropy of the domain wall energy:
\begin{equation}
\Delta B_{2}=B_{2\max }-B_{2\min }=\frac{2}{3}q_{12}^{2}\left\vert f\left(
0\right) -f\left( \pi /4\right) \right\vert =\frac{4}{3}\frac{q_{12}^{2}}{%
\lambda _{1}}\frac{\left\vert \lambda _{1}-\lambda _{2}-2\mu \right\vert }{%
\lambda _{1}+\lambda _{2}+2\mu }.  \label{59a}
\end{equation}%
We found $\Delta B_{2}/B_{2}\sim 4\times 10^{-3}$ for BTO, much smaller
anizotropy $\sim 3\times 10^{-4}$ for PTO, and even smaller one for PZT
where $\Delta B_{2}/B_{2}\sim 4\times 10^{-5}$ (we have used the parameters
listed in the footnote\cite{ElConst}). Such a weak angular dependence of the
domain structure energy is in accordance with phase field results of Ref.%
\cite{Li02} that shows domain walls mainly with thermodynamically favorable
orientations but also with strong deviations from them.

\section{Loss of stability of a single domain state\label{sec:mdloss}}

It is convenient to study the loss of stability of the single domain state
with respect to formation of a domain structure using Eq.~(\ref{eq:Fpa}).
With this, we mean the loss of stability with respect to an arbitrarily
small `polarization waves' so that the original single domain state may be,
in principle, either stable or metastable. Specifically, in our case, when
Eq.~(\ref{23a}) is valid, this state is metastable \cite{BLreview}.

A solution of the equations,
\begin{equation}
\partial \tilde{F}/\partial p=0,\text{ }\partial \tilde{F}/\partial a=0,
\label{60}
\end{equation}%
corresponding to a single domain state ($p\neq 0$, $a=0$) is possible only
if $A_{1}<0$ with $p_{extr}^{2}=-A_{1}/\widetilde{B}$, where the subscript
stands for the `extremum'. This extremum is a minimum (which is relative in
our case) if

\begin{equation}
\partial ^{2}\tilde{F}/\partial p^{2}>0\text{, \ }\partial ^{2}\tilde{F}%
/\partial a^{2}>0,  \label{63}
\end{equation}%
at the point $p=p_{extr}$, $a=0$ given that $\partial ^{2}\tilde{F}/\partial
p\partial a$ is evidently zero at this point. The first inequality in Eq.(%
\ref{63}) is obviously valid for $A_{1}<0$, while the validity of the second
is not immediately evident.

We find from (\ref{eq:Fpa}):%
\begin{equation}
4\left( \frac{\partial ^{2}\tilde{F}}{\partial a^{2}}\right)
_{a=0,p=p_{extr}}=A_{2}+3B_{1}\left( \theta \right)
p_{extr}^{2}=A_{2}-3A_{1}-3A_{1}\left( B_{1}\left( \theta \right) -%
\widetilde{B}\right) /\widetilde{B},  \label{64}
\end{equation}%
From the condition $\left( \partial ^{2}\tilde{F}/\partial a^{2}\right)
_{a=0,p=p_{extr}}=0,$ we obtain the value of $A$ corresponding to a loss of
stability of the single domain state with respect to appearance of a
polarization wave with a given orientation, $A_{pw}\left( \theta \right) $.
It is convenient to present it in the form:%
\begin{equation}
A_{pw}\left( \theta \right) =-\frac{6\pi d}{\varepsilon _{b}d+\epsilon _{e}l}%
+Gk^{2}+\left( \frac{4\pi d}{\varepsilon _{b}d+\epsilon _{e}l}%
-2Gk^{2}\right) \beta \left( \theta \right) ,  \label{65}
\end{equation}%
where%
\begin{equation}
\beta \left( \theta \right) =\frac{q_{11}^{2}-q_{12}^{2}f\left( \theta
\right) \lambda _{1}}{\widetilde{B}\lambda _{1}+2\left[
q_{11}^{2}-q_{12}^{2}f\left( \theta \right) \lambda _{1}\right] }.
\label{65a}
\end{equation}%
The last term in Eq.(\ref{65}) is the result of the polarization-strain
coupling while the first two present the prior case without this coupling
\cite{ChT82,BLreview}. According to Eq.(\ref{63}),\ the corresponding single
domain state will be (meta)stable at low temperatures such that $A<\min
A_{pw}\left( \theta \right) .$
\begin{figure}[tbp]
\begin{center}
\includegraphics[angle=0,width=0.80\textwidth]{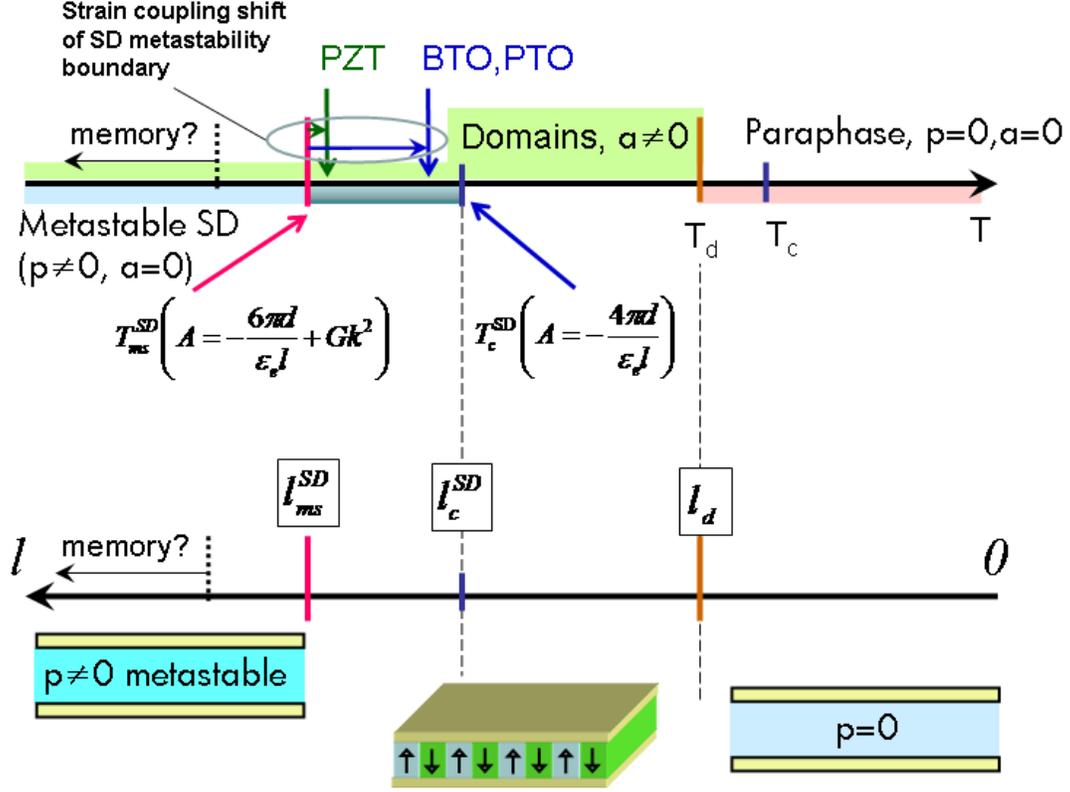}
\end{center}
\caption{(color online) Regions of (meta)stability of the single domain and
polydomain states in the ferroelectric film as a function of temperature $T$
at fixed thickness (top) and as a function of the film thickness $l$ at
fixed $T$ (bottom). Upon lowering the temperature at a fixed thickness $l$
(top panel), the paraphase gives way to domains that are \emph{stable} at
all temperatures $T<T_{d}$, where $T_{d}$ is below the critical temperature
of the bulk ferroelectric transition $T_{c}.$ The single domain (SD)\ state
is \emph{metastable} at low temperatures $T<T_{ms}^{SD}$ , when a strain
coupling is neglected. The strain coupling shifts the boundary of
metastability towards so-called critical temperature for a single domain
state, $T_{c}^{SD},$\ as marked by the vertical arrows for the perovskites
in question. The phase behavior of the films as a function of their
thickness $l$ at fixed temperature (bottom) is qualitatively similar. Very
thin films are in a paraelectric phase that is replaced by domains at larger
thicknesses $l>l_{d}.$ The single domain state is metastable at thicknesses $%
l>l_{ms}^{SD}$ and becomes suitable for memory applications at even larger
(yet to be determined) thicknesses when the life time of the metastable
state becomes sufficiently large. Strain coupling may extend the boundary of
the metastability down to the so-called `critical thickness for SD\
ferroelectric state' $l_{c}^{SD}$. The films with FE memory would be
somewhere at larger thicknesses in the metastable region, as marked on the
diagrams.}
\label{fig:PDiag}
\end{figure}

The actual loss of stability of the single domain state corresponds to the
minimum of $A_{pw}\left( \theta \right) .$ We have seen in Sec.\ref%
{sec:orient} that for perovskites the angular dependencies are very weak and
we can neglect it putting $\ f\lambda _{1}=1.$ Then
\begin{equation}
\beta =\frac{q_{11}^{2}-q_{12}^{2}}{\widetilde{B}\lambda _{1}+2\left(
q_{11}^{2}-q_{12}^{2}\right) }  \label{68}
\end{equation}

Since in BTO, PTO, and PZT $q_{11}^{2}>q_{12}^{2}$ \cite{ElConst}, the last
term in Eq.(\ref{65}) is \emph{positive} and, therefore, the region of \emph{%
meta}stability of the \emph{single} domain state in these systems is \emph{%
broader} than according to \cite{ChT82} and \cite{BLreview}, in apparent
accordance with \cite{perkohl07}. However, there are serious reservations.
First of all, the effect is not very spectacular. Indeed, the factor $\beta $%
\ in the last term of Eq.(\ref{65}) is always less than one half, $\beta
<1/2,$ approaching that value when $q_{11}^{2}-q_{12}^{2}\rightarrow \infty $%
. Therefore,
\begin{equation}
A_{pw}<-\frac{4\pi d}{\varepsilon _{b}d+\epsilon _{e}l},  \label{71}
\end{equation}%
where the r.h.s. corresponds to a \emph{very strong} strain coupling. Recall
that $A=-4\pi d/\left( \varepsilon _{b}d+\epsilon _{e}l\right) $, or $%
A_{1}=0,$ corresponds to what was calculated in several papers as a `\emph{%
critical thickness of single-domain ferroelectricity}' $l_{c}^{SD}$, Eq.(\ref%
{eq:lcSD}).
\begin{figure}[tbp]
\begin{center}
\includegraphics[angle=0,width=0.80\textwidth]{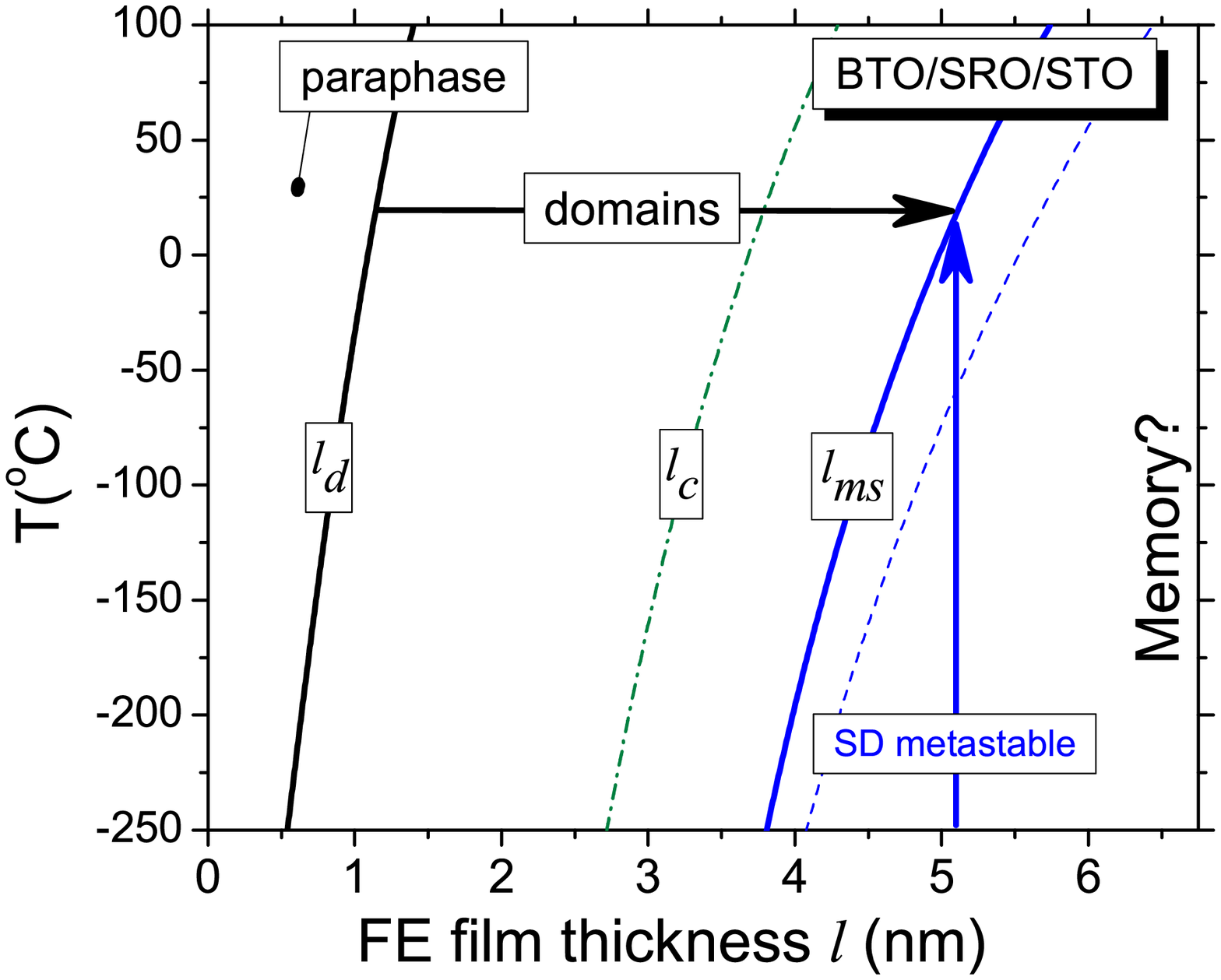}
\end{center}
\caption{ (color online) The phase diagram for BaTiO$_{3}$/SrRuO$_{3}$/SrTiO$%
_{3}$ films in the plane temperature-thickness. The line marked $l_{d}$
delineates the para- and domain phases, while the one marked $l_{ms}$ marks
the boundary of the metastability regions of the single domain state
calculated for BTO accounting for strain coupling, Eq.~(\protect\ref{65}).
The broken line at larger thicknesses from $l_{ms}$ marks the metastable
region calculated without accounting for the strain coupling, Eq.(\protect
\ref{eq:Alms}). The films with FE memory would be somewhere at larger
thicknesses in the metastable region. }
\label{fig:ldlms}
\end{figure}

To get the opposite limit of a \emph{weak strain coupling} for $A_{pw}\left(
0\right) ,$ we put $q_{11}^{2}-q_{12}^{2}=0$ and neglect $Gk^{2},$ as
Pertsev and Kohlstedt \cite{perkohl07} did. We see that
\begin{equation}
-\frac{6\pi d}{\varepsilon _{b}d+\epsilon _{e}l}<A_{pw}\left( 0\right) <-%
\frac{4\pi d}{\varepsilon _{b}d+\epsilon _{e}l},  \label{73}
\end{equation}%
i.e. because of account for the polarization-strain coupling the value of $%
A_{pw}\left( 0\right) $ changes always by less than $1.5$ times. In usual
situation when $\varepsilon _{b}d<\epsilon _{e}l$ this is also the interval
of change of the thickness corresponding to absolute loss of stability of
the single domain state at a \emph{fixed temperature.}

The above moderate, {\normalsize less than} 50\%, range of change is in
striking disagreement with a statement by Pertsev and Kohlstedt \cite%
{perkohl07} who claimed a more than an order of magnitude change by
nullifying the electrostrictive constants. They do not report the details of
their procedure, but it is clear from the rest of the paper that their
suggestion of putting the electrostrictive constants to zero implied changes
in the coefficients of the LGD free energy that should have been
renormalized by the misfit strains. Evidently, it has nothing to do with the
effects of the polarization-strain coupling omitted in Refs \cite%
{ChT82,BLreview}, since this renormalization is automatically taken into
account there, while the effect of misfit strain on LGD coefficients was
apparently neglected in a gedanken exercise performed in Ref.\cite{perkohl07}%
.

Specifically, we find that for the perovskites BTO\ and PTO $\beta =0.4$,
while in PZT\ this parameter is $0.1,$\ i.e. four times smaller. We see that
BTO\ and PTO\ are similar and closer to the limit $q_{11}^{2}-q_{12}^{2}%
\rightarrow \infty ,$ $\beta =0.5$, i.e. the point of loss of stability of
single domain state is quite close in these materials to the 'critical
thickness of single-domain ferroelectricity' studied in Ref.\cite{ghosez03}
and elsewhere. However, the latter does not have any practical importance
because if one fixes the temperature and reduces the film thickness starting
with a monodomain ferroelectric state at low temperatures or large
thicknesses, that state will give way to domains before the thickness
determined by the limit of the single domain state stability is reached. The
fact of the matter is that the single domain state is metastable, it may
have a large lifetime at low temperatures and large film thicknesses but
this lifetime goes essentially to zero (to `atomically' short times) when
the above mentioned temperature or thickness are reached.

Importantly, it follows from Eq.(\ref{65}) that if $q_{11}^{2}<q_{12}^{2},$\
the account for the inhomogeneous strains shrinks the\ region of
metastability of the single domain state. This shows that contrary to the
claim by Pertsev and Kohlstedt there is no general physical phenomenon such
as stabilization of a single domain state because of inhomogeneous\ strains
accompanying formation of domains \ This may seem surprising, because solids
are known to `dislike' the inhomogeneous strains (free energy usually goes
up). Moreover, the expectation of Pertsev and Kohlstedt is justified for a
free-standing film, at least for elastically isotropic solid \cite{BLKhach}.
But it is not certain for a film on substrate considered both by them and in
the present work. To explain the physical reason, we recall that the
coupling with strain renormalizes the coefficients before fourth-order terms
in the LGD free energy, in our case we mean the coefficients before $p^{4}$\
and $p^{2}a^{2}$\ terms. Then, one has to take into account that the
homogeneous strains in the plane of the substrate are not possible while
inhomogeneous ones are. Both homogeneous and inhomogeneous polarization
create homogeneous strain but to a different extent see Eqs.(\ref{32}) and (%
\ref{35}).while inhomogeneous strains are created, of course, by the
inhomogeneous polarization only. The out-of-plane and in-plane strains
couple with the ferroelectric polarization $P_{z}$ by electrostriction terms
with different coefficients and the final result of renormalization of the
coefficient of $p^{2}a^{2}$ term is due to several contributions and it is
not clear upfront. It should be obtained by a consistent analysis, as it has
been done above. No reason is seen\ to discard the possibility that the
inequality $q_{11}^{2}<q_{12}^{2}$ can be realized in some systems, and one
cannot exclude, at least for the moment, the possibility of favoring the
multidomain state by the polarization-strain coupling. Interestingly enough,
this favoring may be very strong: according to Eq.(\ref{65}), the increase
of the region of absolute instability of single domain state becomes
infinite when $q_{12}^{2}-q_{11}^{2}$\ tends to $\widetilde{B}\lambda _{1}/2$
from below.

The `phase diagrams' for the epitaxial FE films on a misfit substrate are
plotted in Figs.~\ref{fig:PDiag}, \ref{fig:ldlms}. The boundaries of
paraelectric phase, domains, and metastable single domain region for BaTiO$%
_{3}$/SrRuO$_{3}$/SrTiO$_{3}$ system are shown in the temperature-film
thickness ($T-l)$ plane in Fig.\ref{fig:ldlms}. They are found from the
conditions that we discussed above and write down here for a reference:

\begin{eqnarray}
A &=&-2Gk^{2}=-4\left( \frac{\pi ^{3}G}{\varepsilon _{\perp }}\right) ^{1/2}%
\frac{1}{l_{d}},  \label{eq:Ald} \\
A &=&-\frac{4\pi d}{\epsilon _{e}l_{c}^{SD}},  \label{eq:AlSD} \\
A &=&-\frac{6\pi d}{\epsilon _{e}l_{ms}^{SD}},  \label{eq:Alms}
\end{eqnarray}%
where the Landau coefficient $A$ is evaluated for a given temperature $T$ of
interest, $d/2=\lambda =0.8\mathring{A},$ $\epsilon _{e}=8.45$ for SrRuO$%
_{3} $ electrode, $G=0.3$\AA $^{2},$ and $\varepsilon _{\perp }$ the
dielectric constant [see its definition below Eq.(\ref{16})] in the plane of
the FE film has been found from the Landau coefficients\cite%
{BLapl06,BLreview}. The last condition corresponds to $l=l_{ms}^{SD}$ found
\emph{without} accounting for the strain coupling. The critical line $%
T-l_{ms}^{SD}$ in the phase diagram, Fig.\ref{fig:ldlms} for BTO \emph{with}
an account for strain coupling has been found from Eq.(\ref{65}). The arrows
on Fig.\ref{fig:ldlms} show the evolution of the state at either $T=\mathrm{%
const}$ or $l=\mathrm{const}$.

One should understand that in both illustrations it is implied that the
corresponding critical points have physical values as solutions to the
conditions (\ref{eq:Tlcond}), or, equivalently, Eqs.(\ref{eq:Td}),(\ref%
{eq:TcSD}),(\ref{eq:ld}),(\ref{eq:lcSD}). Consider first the lowering of the
temperature at a fixed thickness $l$ (Fig.~\ref{fig:PDiag}, top panel),
where the paraphase transforms into domain state below the temperature $%
T_{d} $ that is smaller than the critical temperature of the bulk
ferroelectric transition $T_{c}.$ We see that the single domain (SD)\ state
would be \emph{metastable} at low temperatures $T<T_{ms}^{SD}$ in the region
overlapping with the domain state. Note that the $T_{ms}^{SD}$ plotted in
Fig.\ref{fig:ldlms} is found without accounting for the strain coupling. The
strain coupling then shifts the boundary of metastability towards the
so-called critical temperature for a single domain state $T_{c}^{SD}$\ thus
broadening the range of metastability of the SD state, as shown in Fig.\ref%
{fig:PDiag}. The phase behavior of the films as a function of their
thickness $l$ at fixed temperature (Fig.~\ref{fig:PDiag}, bottom panel) is
qualitatively similar. Very thin films remain in a paraelectric phase that
is replaced by the domains at larger thicknesses $l>l_{d}.$ We see that both
$T_{c}^{SD}$ and $l_{c}^{SD}$ are actually \emph{unreachable} in the present
case, since the system may get to those points only by moving from the
paraphase down (right to left on the phase diagram, Fig.\ref{fig:ldlms}),
but such transitions are preempted by the domain instability that sets in
first. The single domain state is metastable at thicknesses $l>l_{ms}^{SD},$
and becomes suitable for \emph{memory} applications at even larger (yet to
be determined) thicknesses where its life time becomes sufficiently long.

\section{Instability of the checkerboard domain structure\label{sec:checker}}

In the previous Sections, we have assumed that the domain structure is
stripe-like by taking into account only one `polarization wave'. This and
other possibilities have been studied by Chensky and Tarasenko \cite{ChT82}
who considered the uniaxial ferroelectric isotropic in the $x-y$ plane.
Along with the stripe-like structure they discussed also the checkerboard
and the hexagonal domain structures. The latter can be realized in the
presence of an external field only which is not discussed in this paper.
However, a checkerboard structure should be analyzed as an alternative to
the stripe structure. In Ref.\cite{ChT82}, the authors stated that the
checkerboard structure never realizes, although, surprisingly, there is no
proof of this statement. In this Section, we shall\ show that this structure
is indeed unstable for the isotropic case treated in Ref.\cite{ChT82} and
then show that this conclusion holds also when one explicitly takes into
account the \emph{polarization-strain interaction}, apart from mere
renormalization of the LGD coefficients by the misfit strains.

Once again, we consider a short-circuited sample, i.e.the ferroelectric
polarization is described by:%
\begin{equation}
P_{z}=a_{1}\cos \boldsymbol{k}_{1}\boldsymbol{r}\cos qz+a_{2}\cos
\boldsymbol{k}_{2}\boldsymbol{r}\cos qz,  \label{79}
\end{equation}%
where $\boldsymbol{k}_{1}$ and $\boldsymbol{k}_{2}$ are two noncollinear
vectors whose modulus is given by Eq.(\ref{eq:k}) and whose (mutually
orthogonal) directions remain unspecified for a moment, Fig.~\ref%
{fig:checker}.
\begin{figure}[tbp]
\begin{center}
\includegraphics[angle=0,width=0.60\textwidth]{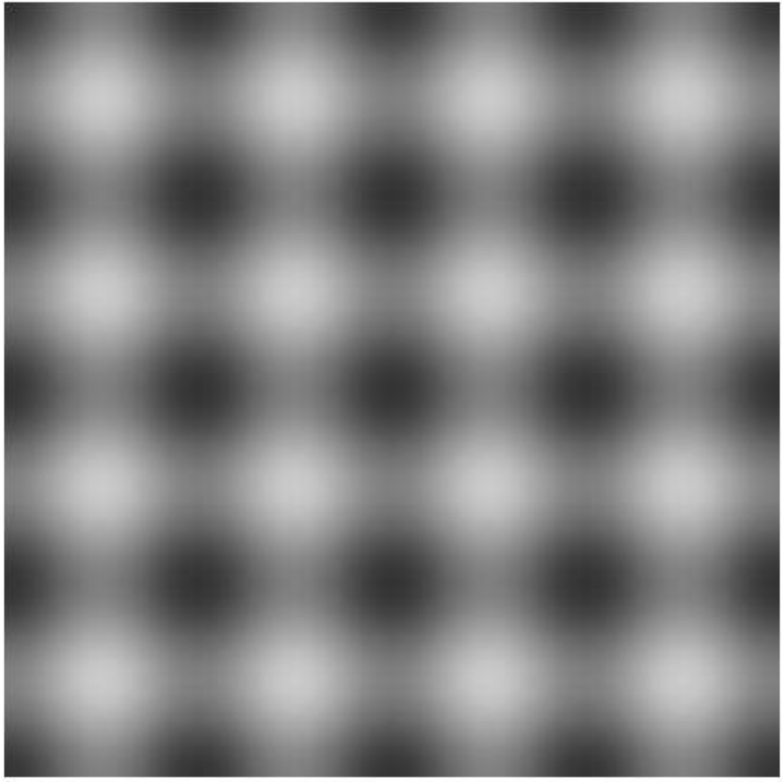}
\end{center}
\caption{ Schematic of the checkerboard domain structure. It is absolutely
unstable in case of BaTiO$_{3}$, PbTiO$_{3}$, and Pb(Zr$_{0.5}$Ti$_{0.5}$)O$%
_{3}$ typical perovskite ferroelectrics.}
\label{fig:checker}
\end{figure}

\subsection{Checkerboard domains \emph{without} elastic coupling ($%
q_{11}=q_{12}=0)$}

In this case the solution for the fields is [cf. Eq.(\ref{12})]{\LARGE \ }%
\begin{eqnarray}
E_{z} &=&E_{z}^{k1}\cos \boldsymbol{k}_{1}\boldsymbol{r}\cos
qz+E_{z}^{k2}\cos \boldsymbol{k}_{2}\boldsymbol{r}\cos qz, \\
E_{x,y} &=&E_{x,y}^{k1}\sin \boldsymbol{k}_{1}\boldsymbol{r}\sin
qz+E_{x,y}^{k2}\sin \boldsymbol{k}_{2}\boldsymbol{r}\sin qz,
\end{eqnarray}%
with Eq.(\ref{19}) still applicable to spatial harmonics (as follows from
linearity of Maxwell equations) and the equation of state for the
fundamental harmonics is the same as Eq.(\ref{14}):
\begin{equation}
A_{2}a_{1}+\left[ BP_{z}^{3}\right] _{\mathrm{cc}}=E_{z}^{k1},
\end{equation}%
where one retains the terms $\left[ BP_{z}^{3}\right] _{\mathrm{cc}}\propto
\cos \boldsymbol{k}_{1}\boldsymbol{r}\cos qz$ (symmetry dictates the
analogous expressions for $a_{2}$). In the above equation,
\begin{equation}
P_{z}^{3}=\left( a_{1}\cos \boldsymbol{k}_{1}\boldsymbol{r}+a_{2}\cos
\boldsymbol{k}_{2}\boldsymbol{r}\right) ^{3}\cos ^{3}qz=\frac{9}{16}\left(
a_{1}^{3}+2a_{1}a_{2}^{2}\right) \cos \boldsymbol{k}_{1}\boldsymbol{r}\cos
qz+\frac{9}{16}\left( a_{2}^{3}+2a_{2}a_{1}^{2}\right) \cos \boldsymbol{k}%
_{2}\boldsymbol{r}\cos qz+\ldots ,
\end{equation}%
so that we obtain for the fundamental harmonic the following equations of
state:%
\begin{equation}
A_{2}a_{1}+\frac{9B}{16}\left( a_{1}^{3}+2a_{1}a_{2}^{2}\right) =0
\label{eq:a1a2noel}
\end{equation}%
and the analogous equation for $a_{2}$. Since these equations are obtained
from extremum of the free energy, $\partial \widetilde{F}/\partial
a_{1(2)}=0,$ we again restore the full free energy, accounting for the
symmetric contribution by the $a_{2}$ harmonic:%
\begin{equation}
V^{-1}\widetilde{F}=\frac{A_{2}}{8}(a_{1}^{2}+a_{2}^{2})+\frac{9}{256}%
B\left( a_{1}^{4}+a_{2}^{4}\right) +\frac{9}{64}Ba_{1}^{2}a_{2}^{2}.
\label{eq:chk0}
\end{equation}%
The equations of state have the checkerboard solution:
\begin{equation}
a_{1}^{2}=a_{2}^{2}=-16A_{2}/27B.
\end{equation}%
Checking what type of extremum for the free energy is this solution,%
\begin{equation*}
\frac{\partial ^{2}F}{\partial a_{1}^{2}}\times \frac{\partial ^{2}F}{%
\partial a_{2}^{2}}-\left( \frac{\partial ^{2}F}{\partial a_{1}\partial a_{2}%
}\right) ^{2}=-\frac{1}{12}A_{2}^{2}<0,
\end{equation*}%
we see that the checkerboard solution is the \emph{maximum} of the free
energy for some directions in the $a_{1},a_{2}$\ plane and is \emph{%
absolutely unstable}.

\subsection{Checkerboard domains \emph{with} elastic coupling}

We have seen above that the elastic coupling renormalizes the fourth order
coefficients in formulas like Eq. (\ref{eq:chk0}) reducing them by some
amounts which are different for different coefficients. Thus, instead of Eq.(%
\ref{eq:chk0}), we will have:%
\begin{equation}
V^{-1}\widetilde{F}=\frac{A_{2}}{8}(a_{1}^{2}+a_{2}^{2})+\frac{9}{256}\left(
B_{21}a_{1}^{4}+B_{22}a_{2}^{4}\right) +\frac{9}{64}B_{3}a_{1}^{2}a_{2}^{2},
\label{checkerfreeen}
\end{equation}%
where $B_{21}$ and $B_{22}$ are given by Eq.(\ref{49}) for the corresponding
angles and $B_{3}$ is a new coefficient which depends on the both angles and
which can be, in principle, both positive and negative. Both from Eq.(\ref%
{49}) and the cubic symmetry one realizes that $B_{21}=$ $B_{22}=B_{2}$. For
what follows, it is important to mention that when $B_{3}$ is negative it
cannot be of large absolute value, otherwise there will be directions in the
$(a_{1},a_{2})$ plane along which the free energy diminishes without limits
at large values of $a_{1},a_{2}$ what means a global instability of the
system. By putting $a_{1}=a_{2}$, one sees from Eq. (\ref{checkerfreeen})
that to avoid this instability the condition
\begin{equation}
B_{2}+2B_{3}>0  \label{cond}
\end{equation}%
should be fulfilled. Another evident condition of the global stability is $%
B_{2}>0$.

The checkerboard solution is:
\begin{equation}
a_{1}^{2}=a_{2}^{2}=\frac{-16A_{2}}{9\left( B_{2}+2B_{3}\right) }.
\end{equation}%
To analyze stability of this solution, we calculate the second derivatives%
\begin{equation*}
\frac{\partial ^{2}F}{\partial a_{1}^{2}}=\frac{\partial ^{2}F}{\partial
a_{2}^{2}}=\frac{A_{2}}{4}+\frac{27}{64}B_{2}a_{1(2)}^{2}+\frac{9}{32}%
B_{3}a_{2(1)}^{2}=-\frac{1}{2}A_{2}\frac{B_{2}}{B_{2}+2B_{3}},
\end{equation*}%
\begin{equation*}
\frac{\partial ^{2}F}{\partial a_{1}\partial a_{2}}=\frac{9}{16}%
B_{3}a_{1}a_{2}=\pm \frac{A_{2}B_{3}}{B_{2}+2B_{3}}
\end{equation*}%
we then find the discriminant

\begin{equation}
Z\equiv \left( \frac{\partial ^{2}F}{\partial a_{1}^{2}}\right) \left( \frac{%
\partial ^{2}F}{\partial a_{2}^{2}}\right) -\left( \frac{\partial ^{2}F}{%
\partial a_{1}\partial a_{2}}\right) ^{2}=\frac{1}{4}A_{2}^{2}\frac{%
B_{2}-2B_{3}}{B_{2}+2B_{3}}.  \label{eq:F"checker}
\end{equation}%
Our further study is aimed at finding out if and when the condition of
positiveness of $Z$, i.e.$B_{2}>2B_{3}$, is compatible with the two
conditions of the global stability mentioned above. Thus we need formulas
for the coefficients $B_{2}$ and $B_{3}$.

Turning to taking into account explicitly the polarization-strain coupling,
we recall that in our approximation of sufficiently thick film the only
source of the elastic strains is $P_{z}^{2}.$ This function contains now a
cross term stemming from

\begin{eqnarray}
P_{z}^{2} &=&(a_{1}\cos \boldsymbol{k}_{1}\boldsymbol{r}\cos qz+a_{2}\cos
\boldsymbol{k}_{2}\boldsymbol{r}\cos qz)^{2} \\
&=&\left[ a_{1}^{2}+a_{2}^{2}+2a_{1}a_{2}\left( \cos \boldsymbol{p}^{+}%
\boldsymbol{r}+\cos \boldsymbol{p}^{-}\boldsymbol{r}\right) +a_{1}^{2}\cos
2k_{1}r+a_{2}^{2}\cos 2k_{2}r\right] \frac{1+\cos 2qz}{4},
\end{eqnarray}%
where $\boldsymbol{p}^{\pm }=\boldsymbol{k}_{1}\pm \boldsymbol{k}_{2}$.
Naturally, the components of the strain tensor will have terms depending on $%
\cos \boldsymbol{p}^{+}\boldsymbol{r}$, $\cos \boldsymbol{p}^{-}\boldsymbol{r%
}$:. We will have for $u_{xx}$%
\begin{eqnarray*}
u_{xx} &=&u_{xx}^{(0)}+u_{xx}^{(1)}\cos 2qz+u_{xx}^{p+}\cos \boldsymbol{p}%
^{+}\boldsymbol{r}+u_{xx}^{p-}\cos \boldsymbol{p}^{-}\boldsymbol{r}%
+u_{xx}^{q+}\cos \boldsymbol{p}^{+}\boldsymbol{r}\cos 2qz \\
&&+u_{xx}^{q-}\cos \boldsymbol{p}^{-}\boldsymbol{r}\cos
2qz+u_{1,xx}^{(3)}\cos 2k_{1}r+u_{2,xx}^{(3)}\cos 2k_{2}r+\left(
u_{1,xx}^{(4)}\cos 2k_{1}r+u_{2,xx}^{(4)}\cos 2k_{2}r\right) \cos 2qz,
\end{eqnarray*}%
and the analogous equations for $u_{yy}$ and $u_{zz}.$

From our previous experience, it becomes immediately clear that $%
u_{xx(yy)}^{p\pm }=u_{xx(yy)}^{q\pm }$, since the equations for those
components do not depend on $z$ and we can now drop the indices $p$ and $q$
from the corresponding terms, leaving only $u_{xx(yy)}^{+,-}.$ Also, due to
the same reason as above, $u_{xx\left( yy\right) }^{(0)}=u_{xx\left(
yy\right) }^{(1)}=0$. Then, one can write:%
\begin{eqnarray*}
u_{xx} &=&\left( u_{xx}^{+}\cos \boldsymbol{p}^{+}\boldsymbol{r}%
+u_{xx}^{-}\cos \boldsymbol{p}^{-}\boldsymbol{r}\right) (1+\cos 2qz) \\
&&+u_{1,xx}^{(3)}\cos 2\boldsymbol{k}_{1}\boldsymbol{r}+u_{2,xx}^{(3)}\cos 2%
\boldsymbol{k}_{2}\boldsymbol{r}+(u_{1,xx}^{(4)}\cos 2\boldsymbol{k}_{1}%
\boldsymbol{r}+u_{2,xx}^{(4)}\cos 2\boldsymbol{k}_{2}\boldsymbol{r})\cos 2qz.
\end{eqnarray*}%
and a similar equation for $u_{yy}$. The `diagonal' terms for the first
(second) $\boldsymbol{k}_{1(2)}$ harmonic\ are
\begin{equation*}
u_{1(2),xx}^{(3)}+u_{1(2),yy}^{(3)}=u_{1(2),xx}^{(4)}+u_{1(2),yy}^{(4)}=-q_{12}%
\frac{a_{1\left( 2\right) }^{2}f\left( \theta _{1\left( 2\right) }\right) }{4%
}.
\end{equation*}

All the cross terms satisfy

\begin{equation}
\lambda _{1}\frac{\partial ^{2}u_{x}^{\pm }}{\partial x^{2}}+(\lambda
_{2}+\mu )\frac{\partial ^{2}u_{y}^{\pm }}{\partial y\partial x}+\mu \frac{%
\partial ^{2}u_{x}^{\pm }}{\partial y^{2}}+q_{12}\frac{\partial P_{z}^{2\pm }%
}{\partial x}=0,
\end{equation}%
\begin{equation}
\lambda _{1}\frac{\partial ^{2}u_{y}^{\pm }}{\partial y^{2}}+(\lambda
_{2}+\mu )\frac{\partial ^{2}u_{x}^{\pm }}{\partial y\partial x}+\mu \frac{%
\partial ^{2}u_{y}^{\pm }}{\partial x^{2}}+q_{12}\frac{\partial P_{z}^{2\pm }%
}{\partial y}=0,
\end{equation}%
where $P_{z}^{2\pm }=\frac{1}{2}a_{1}a_{2}\cos \boldsymbol{p}^{\pm }%
\boldsymbol{r},$ in components $u_{x(y)}^{\pm }\propto \sin \boldsymbol{p}%
^{\pm }\boldsymbol{r},$ $\partial ^{2}u_{i}^{\pm }/\partial x^{2}=-\left(
p_{x}^{\pm }\right) ^{2}u_{i}^{\pm }$ etc.

\begin{equation}
\lambda _{1}p_{x}^{\pm 2}u_{x}^{\pm }+(\lambda _{2}+\mu )p_{x}^{\pm
}p_{y}^{\pm }u_{y}^{\pm }+\mu p_{y}^{\pm 2}u_{x}^{\pm }+q_{12}p_{x}^{\pm }%
\frac{1}{2}a_{1}a_{2}=0,
\end{equation}%
\begin{equation}
\lambda _{1}p_{y}^{\pm 2}u_{y}^{\pm }+\left( \lambda _{2}+\mu \right)
p_{x}^{\pm }p_{y}^{\pm }u_{x}^{\pm }+\mu p_{x}^{\pm 2}u_{y}^{\pm
}+q_{12}p_{y}^{\pm }\frac{1}{2}a_{1}a_{2}=0.
\end{equation}%
Then,
\begin{equation}
u_{xx}^{\pm }+u_{yy}^{\pm }=p_{x}^{\pm }u_{x}^{\pm }+p_{y}^{\pm }u_{y}^{\pm
}=-q_{12}\frac{a_{1}a_{2}}{2}f(\theta ^{\pm }).  \label{eq:dilpm}
\end{equation}%
For $u_{zz}$ we conclude, as above, that only $u_{zz}^{(0)}$and $%
u_{zz}^{(1)} $ are non-zero and following the same reasoning as for the
stripe phase, we obtain%
\begin{equation*}
u_{zz}^{(0)}=u_{zz}^{(1)}=-\frac{q_{11}}{4\lambda _{1}}(a_{1}^{2}+a_{2}^{2}).
\end{equation*}%
Having solved the elastic problem, we are now in a position to write down
the constituent equations containing $a_{1}$ and $a_{2}$ only. To this end,
we write two equations for $a_{1}$ and $a_{2}$ analogous to Eq.(\ref{21})
but this time $\left[ \ldots \right] _{\mathrm{cc}}$ would mean the
proportionality to $\cos \boldsymbol{k}_{1}\boldsymbol{r}\cos qz$ or $\cos
\boldsymbol{k}_{2}\boldsymbol{r}\cos qz,$ respectively. Since both equations
have the same structure, we will discuss that for $a_{1}$ only and, for the
sake of brevity, we will mention only the terms containing $a_{2}$, the
other terms are the same as for the one-sinusoid case discussed above.

For clarity sake, we repeat Eq.(\ref{21}) with a minor change for the
present case:
\begin{equation}
A_{2}a_{1}+\left[ BP_{z}^{3}+2q_{11}P_{z}u_{zz}+2q_{12}P_{z}\left(
u_{xx}+u_{yy}\right) \right] _{\mathrm{cc}}=0.
\end{equation}%
It is straightforward to find that the $a_{2}$-containing term stemming from
$\left[ P_{z}^{3}\right] _{cc}$ is $9a_{1}a_{2}^{2}/8.$ \ From Eq.(\ref{31}%
), one sees that $\left[ P_{z}u_{zz}\right] _{cc}=a_{1}\left( u_{zz}^{(0)}+%
\frac{1}{2}u_{zz}^{(1)}\right) $, recall that now we consider the case $p=0$%
, and the contribution of this term is
\begin{equation*}
-\frac{3q_{11}}{8\lambda _{1}}a_{1}a_{2}^{2}.
\end{equation*}%
Now%
\begin{equation*}
\left[ P_{z}\left( u_{xx}+u_{yy}\right) \right] _{cc}=\frac{a_{1}}{2}\left(
u_{1,xx}^{(3)}+u_{1,yy}^{(3)}\right) +\frac{a_{1}}{4}\left(
u_{1,xx}^{(4)}+u_{1,yy}^{(4)}\right) +\frac{3}{4}%
a_{2}(u_{xx}^{+}+u_{yy}^{+}+u_{xx}^{-}+u_{yy}^{-})
\end{equation*}%
and contribution of this term to the equation of state is%
\begin{equation*}
-q_{12}\frac{3a_{1}a_{2}}{8}\left[ f(\theta ^{+})+f(\theta ^{-})\right] .
\end{equation*}%
Finally, the constituent equation for $a_{1}$ takes the form:%
\begin{equation*}
A_{2}a_{1}+\frac{9}{16}a_{1}^{3}B_{2}\left( \theta _{1}\right) +\frac{9}{8}%
a_{1}a_{2}^{2}B_{3}\left( \theta ^{+},\theta ^{-}\right) =0,
\end{equation*}%
where we have introduced
\begin{equation}
B_{3}\left( \theta ^{+},\theta ^{-}\right) =B-\frac{2}{3}\frac{q_{11}^{2}}{%
\lambda _{1}}-\frac{2}{3}q_{12}^{2}\left[ f(\theta ^{+})+f(\theta ^{-})%
\right]  \label{B3}
\end{equation}%
Similarly to the case of one sinusoid, we recover the free energy:
\begin{equation}
F=\frac{A_{2}}{8}\left( a_{1}^{2.}+a_{2}^{2}\right) +\frac{9}{256}%
B_{2}\left( \theta _{1}\right) a_{1}^{4}+\frac{9}{256}B_{2}\left( \theta
_{2}\right) a_{2}^{4}+\frac{9}{64}B_{3}\left( \theta ^{+},\theta ^{-}\right)
a_{1}^{2}a_{2}^{2},  \label{checkfreeen}
\end{equation}%
Recall that the square symmetry suggests that $B_{2}\left( \theta
_{1}\right) =B_{2}\left( \theta _{2}\right) $. and $B_{2}\left( \theta
\right) $\ is given by Eq.~(\ref{49})

Turning to examining the sign of the discriminant $Z$, we should mention
that according to Eqs.(\ref{49}) and (\ref{B3}):
\begin{equation*}
B_{2}-2B_{3}=-B+\frac{4}{3}q_{12}^{2}\left[ f(\theta ^{+})+f(\theta ^{-})-%
\frac{1}{2}f\left( \theta _{1}\right) \right] .
\end{equation*}%
One sees that the maxima of $Z$ correspond to maxima of $f(\theta ^{+})$ and
$f(\theta ^{-})$ [note that $f(\theta _{\max }^{+})=f(\theta _{\max }^{-})$
because of the cubic symmetry], which are, automatically, the minima of $%
f\left( \theta _{1}\right) $ as we have seen in Sec.\ref{sec:orient}. Then,%
\begin{equation*}
\left[ B_{2}-2B_{3}\right] _{\max }=-B+\frac{4}{3}q_{12}^{2}\left[ 2f(\theta
_{\max })-\frac{1}{2}f\left( \theta _{\min }\right) \right] .
\end{equation*}%
Using the values of $f(\theta _{\max })$ and $f\left( \theta _{\min }\right)
$ found in Sec.\ref{sec:orient}, we find that if $\lambda _{2}+2\mu >$ $%
\lambda _{1},$%
\begin{equation}
\left[ B_{2}-2B_{3}\right] _{\max }=-B+\frac{4}{3}q_{12}^{2}\left( \frac{2}{%
\lambda _{1}}-\frac{1}{\lambda _{1}+\lambda _{2}+2\mu }\right) =-B+\frac{4}{3%
}q_{12}^{2}\frac{2\left( \lambda _{2}+2\mu \right) +\lambda _{1}}{\lambda
_{1}\left( \lambda _{1}+\lambda _{2}+2\mu \right) }  \label{stab1}
\end{equation}%
and if $\lambda _{2}+2\mu <$ $\lambda _{1},$%
\begin{equation}
\left[ B_{2}-2B_{3}\right] _{\max }=-B+\frac{4}{3}q_{12}^{2}\left( \frac{4}{%
\lambda _{1}+\lambda _{2}+2\mu }-\frac{1}{2\lambda _{1}}\right) =-B+\frac{4}{%
3}q_{12}^{2}\frac{7\lambda _{1}-\lambda _{2}-2\mu }{2\lambda _{1}\left(
\lambda _{1}+\lambda _{2}+2\mu \right) }  \label{stab2}
\end{equation}

To prove that the checkerboard structure can be stable, at least in
principle, with respect to small fluctuations, we should demonstrate that
the positiveness of $\left[ B_{2}-2B_{3}\right] _{\max }$ is compatible with
the conditions $B_{2}>0$ and $\left[ B_{2}-2B_{3}\right] _{\min }>0$ which
guarantee the global stability of the system. We do not intend to perform an
exhaustive analysis but want only to demonstrate that this is possible under
certain conditions, unlike in the case without the elastic coupling. As an
example, we consider a system with a weak elastic anisotropy which is valid
for the perovskites, i.e. we shall assume $\lambda _{2}+2\mu $ $\simeq
\lambda _{1}$, and $q_{11}=0.$ Both Eq.(\ref{stab1}) and Eq.(\ref{stab2}) \
then give
\begin{equation}
\left[ B_{2}-2B_{3}\right] _{\max }\simeq -B+\frac{2q_{12}^{2}}{\lambda _{1}}%
,  \label{stab3}
\end{equation}%
and give for the positiveness of $\left[ B_{2}-2B_{3}\right] _{\max }$ \ the
same condition:$q_{12}^{2}>B\lambda _{1}/2$ while the condition $B_{2}>0$
now reads $q_{12}^{2}<3B\lambda _{1}/2.$ One sees that for a nearly
elastically isotropic ferroelectric with $q_{11}=0,$ the checkerboard
structure is at least metastable if
\begin{equation*}
3B\lambda _{1}/2>q_{12}^{2}>B\lambda _{1}/2
\end{equation*}%
Of course, the above set of the material coefficients looks fairly exotic
but it is just an example aimed at nothing more but demonstration that the
checkerboard domain structures are permitted due to the elastic coupling
when certain conditions on the material coefficients are met.

In case of real perovskite films the checkerboard structure is not stable.
To see this, we can rewrite Eq.~(\ref{stab3}) in the form:%
\begin{equation*}
\left[ B_{2}-2B_{3}\right] _{\max }\simeq -\widetilde{B}-\frac{2}{\lambda
_{1}}\left( q_{11}^{2}-q_{12}^{2}\right) <0.
\end{equation*}%
Indeed, we have already mentioned above that for the perovskites $%
q_{11}^{2}>q_{12}^{2}$. Also, $\widetilde{B}>0$ there. Therefore, in the
perovskites the checkerboard domain structure is \emph{absolutely unstable}

\section{Conclusions\label{sec:summ}}

With the use of the Landau-Ginzburg-Devonshire theory, we have studied the
effects of polarization-strain coupling when defining the character of
equilibrium domain structures and the limits of absolute instability of a
single domain state in thin films of cubic ferroelectric films on a misfit
substrate. On the compressive substrate, the cubic ferroelectric behaves
substantially as a uniaxial ferroelectric with the polar axis perpendicular
to the film. The film is sandwiched between the electrodes that do not
provide a perfect screening of the depolarizing field because of the finite
Thomas-Fermi screening.length. Such a system is exemplified by (100) BaTiO$%
_{3}$/SrRuO$_{3}$/SrTiO$_{3}$ film and similar perovskite structures.
Quantitative results have been obtained for BaTiO$_{3}$, PbTiO$_{3}$, and
Pb(Zr$_{0.5}$Ti$_{0.5})$O$_{3}$. We have found that close to the
paraelectric-ferroelectric phase transition or at the film thicknesses close
to the minimal thickness compatible with the ferroelectricity, the
equilibrium domain structure in perovskites is the stripe-wise one with the
stripes parallel (perpendicular) to the cubic axes in BaTiO$_{3}$, PbTiO$%
_{3} $, while running at 45$^{\circ }$ to cubic axes in Pb(Zr$_{0.5}$Ti$%
_{0.5}$)O$_{3}$. The energy of the domain structure depends very weakly on
the stripe orientation, the maximum change proves to be well below 1\% in
all three cases. We found that because of the polarization-strain coupling a
competing checkerboard domain structure may, at least in principle, be
equilibrium or metastable when certain conditions on the material constants
are met, but we are not aware of any material system with such conditions.
The limit of absolute instability of the single domain state changes due to
the polarization-strain coupling. Thus, the interval where the absolute
instability is absent, meaning a metastability in the cases at hand, widens
in perovskites in agreement with the earlier conclusion by Pertsev and
Kohlstedt\cite{perkohl07}. However, this effect is much smaller than that
claimed by them. Increase of the metastability range is substantial in BaTiO$%
_{3}$ and PbTiO$_{3}$ where the absolute instability limit becomes close to
what is often called the "critical thickness for ferroelectricity" $%
l_{c}^{SD},$\cite{ghosez03}\ Fig.\ref{fig:ldlms}, but without accounting for
the domain formation, which sets in first and prevents the system from ever
reaching this point. The effect is much smaller in Pb(Zr$_{0.5}$Ti$_{0.5})$O$%
_{3}$. We have found also that the polarization-strain coupling can lead to
narrowing of the region of relative stability of the single domain state
under certain conditions on the material constants, but we are not aware of
an experimental.realization of these conditions.

APL has been partially supported by Ministry of Science and Education of
Russian Federation (State Contract \# 02.740.11.5156).

\end{document}